\documentclass[amsmath,amssymb,twocolumn,aps,prx,nofootinbib,longbibliography,floatfix]{revtex4-1}
\usepackage{color}
\definecolor{labelkey}{cmyk}{.4,.2,0,0}

\usepackage[caption=false]{subfig}

\usepackage[colorlinks=true,linkcolor=blue,urlcolor=blue,citecolor=blue]{hyperref}

\usepackage{times}

\usepackage[colorinlistoftodos]{todonotes}

\usepackage[normalem]{ulem}

\usepackage{tikz}
\usetikzlibrary{arrows,shapes,positioning}
\usetikzlibrary{decorations.markings}
\tikzstyle arrowstyle=[scale=1]
\tikzstyle directed=[postaction={decorate,decoration={markings,
    mark=at position .65 with {\arrow[arrowstyle]{stealth}}}}]
\tikzstyle endreversedirected=[postaction={decorate,decoration={markings,
    mark=at position 1.0 with {\arrow[arrowstyle]{stealth}}}}]
\tikzstyle enddirected=[postaction={decorate,decoration={markings,
    mark=at position 1.0 with {\arrow[arrowstyle]{stealth}}}}]

\tikzstyle reverse directed=[postaction={decorate,decoration={markings,
    mark=at position .65 with {\arrowreversed[arrowstyle]{stealth};}}}]

\newcommand{\w}{\nn\\&&}
\newcommand{\Mathematica}[1]{}

\newcommand{\suppress}[1]{}
\newcommand{\Eq}[1]{Eq.~(\ref{#1})}
\newcommand{\eq}[1]{(\ref{#1})}

\newcommand{\half}{\frac12}

\newcommand{\bea}{\begin{eqnarray}}
\newcommand{\eea}{\end{eqnarray}}
\newcommand{\beq}{\begin{equation}}
\newcommand{\eeq}{\end{equation}}
\newcommand{\be}{\begin{equation}}
\newcommand{\ee}{\end{equation}}

\newcommand{\rme}{\mathrm{e}}

\newcommand{\nn}{\nonumber}

\newcommand{\nott}[1]{}
\newcommand{\Fig}[1]{\includegraphics[width=\columnwidth]{./#1}} 
\newcommand{\fig}[2]{\includegraphics[width=#1\columnwidth]{./#2}}

\newlength{\bilderlength}

\newcommand{\ca}[1]{\mathcal #1}

\renewcommand{\paragraph}{\subsubsection*}

\providecommand{\bottomrule}{\botrule}
\providecommand{\midrule}{\colrule}

\begin{document}

\bibliographystyle{KAY-hyper}

\title{Fractal dimension of critical curves in the $O(n)$-symmetric   $\phi^4$-model and crossover exponent at 6-loop order: Loop-erased random walks, self-avoiding walks, Ising,  XY and Heisenberg models}
\author{Mikhail Kompaniets${}^{{1}}$ and Kay J\"org Wiese${}^2$}
\affiliation{${^{1}}$ Saint Petersburg State University, 7/9 Universitetskaya nab., Saint Petersburg 199034, Russia.\\
${^{2}}$\mbox{CNRS-Laboratoire de Physique  de l'Ecole Normale
  Sup\'erieure,  ENS, Universit\'e PSL,} Sorbonne Universit\'e, Universit\'e Paris-Diderot, Sorbonne Paris Cit\'e, 24 rue Lhomond, 75005 Paris, France.}

\begin{abstract}
We calculate the fractal dimension $d_{\rm f}$ of critical curves in the  $O(n)$ symmetric $(\vec \phi^2)^2$-theory in $d=4-\epsilon$ dimensions at 6-loop order. This gives the fractal dimension of loop-erased random walks  at $n=-2$, self-avoiding walks ($n=0$), Ising lines $(n=1)$, and XY lines ($n=2$), in agreement with numerical simulations. It can be compared to the fractal   dimension $d_{\rm f}^{\rm tot}$   of all lines, i.e.\ backbone plus the surrounding  loops, identical to $d_{\rm f}^{\rm tot} = 1/\nu$. The combination $\phi_{\rm c}= d_{\rm f}/d_{\rm f}^{\rm tot} = \nu d_{\rm f}$ is the  crossover exponent, describing a system with mass anisotropy. 
Introducing a novel self-consistent resummation procedure, and combining it with analytic results in $d=2$ allows us to give improved estimates in $d=3$ for all relevant exponents at 6-loop order.  
\end{abstract}

\maketitle

\section{Introduction and Summary}
Critical exponents for the $O(n)$-model have been calculated for many years, using high-temperature series expansions \cite{ItzyksonDrouffe1,ItzyksonDrouffe2,ArisueFujiwara2003,ArisueFujiwaraTabata2004,ButeraComi1995,ButeraComi2000,ButeraComi1999}, an expansion\footnote{In this paper we use $d=4-\epsilon$ which is more common for statistical physics, while the original six-loop calculations \cite{BatkovichChetyrkinKompaniets2016,KompanietsPanzer:LL2016,KompanietsPanzer2017} were performed in space dimension $d=4-2\varepsilon$ which is used in high-energy physics.} in $d=4-\epsilon$, \cite{Amit,Zinn,Vasilev2004,ChetyrkinKataevTkachov1981,ChetyrkinKataevTkachov1981b,ChetyrkinGorishnyLarinTkachov1983,
Kazakov1983,KleinertNeuSchulte-FrohlindeChetyrkinLarin1991, BatkovichChetyrkinKompaniets2016,KompanietsPanzer:LL2016,KompanietsPanzer2017,Schnetz2018},   field theory in   dimension $d=3$ \cite{ParisiBook,BakerNickelGreenMeiron1976,GuidaZinn-Justin1998},  Monte Carlo simulations~\cite{Clisby2017,ClisbyDunweg2016,
CampostriniHasenbuschPelissettoVicari2006,HasenbuschVicari2011,Deng2006,
HasenbuschVicari2011,HasenbuschVicari2011},   exact results   in dimension $d=2$ \cite{Nienhuis1987,HenkelCFT,DiFrancescoMathieuSenechal,NishimoriOrtiz2011}, or the conformal bootstrap \cite{PolandRychkovVichi2019,KosPolandSimmons-DuffinVichi2016,El-ShowkPaulosPolandRychkovSimmons-DuffinVichi2014,
Castedo-Echeverrivon-HarlingSerone2016}. Most of these methods rely on some resummation procedure \cite{GuttmanInDombLebowitz13,Zinn-Justin2001,KompanietsPanzer2017}.
The main exponents   are the decay of the 2-point function at $T_{c}$
\be
\left< \phi(x) \phi(0)\right> \sim |x|^{{2-d-\eta}} \ ,
\ee
and 
the divergence of the correlation length $\xi$ as a  function of $T-T_{c}$
\be
\xi \sim |T-T_{c}|^{-\nu}\ .
\ee
Other exponents are related to these \cite{Amit}, as the divergence of the specific heat
\be 
c  \sim |T-T_{c}|^{-\alpha }\ , \qquad \alpha = 2 -\nu d\ , 
\ee
the magnetization $M$ below $T_{c}$
\be
M\sim (T_{c}-T)^{\beta}\ , \qquad\beta = \frac \nu 2 (d-2 +\eta )\ ,
\ee
the susceptibility $\chi$, 
\be
\chi \sim |T-T_{c}|^{-\gamma}\ , \qquad \gamma = \nu (2-\eta)\ , 
\ee
and the magnetization at $T_{c}$ in presence of a magnetic field $h$, 
\be
M \sim h^{  1/\delta}\ , \qquad \delta = \frac{d+2-\eta }{d-2+\eta}\ .
\ee
The renormalization-group treatment starts from the $\phi^{4}$ theory with $O(n)$ symmetry, 
\be\label{S-bare}
{\cal S} = \int_{x} \,  \frac{m_{0}^{2}}2 \vec \phi_{0}(x)^{2} + \frac{1}{2} [\nabla \vec \phi_{0}(x)]^{2} +    g_{0} \frac{16\pi^{2}}{4!}\left[\vec \phi_{0}(x)^{2} \right]^{2}\ , 
\ee
where
$\vec \phi_{0}(x) \in \mathbb{R}^{n}$. 
The index 0 indicates bare quantities. The renormalized action is
\be
{\cal S} =  \int\limits_x \,  Z_{1}\frac{m^{2}}2 \vec \phi(x)^{2} +\frac{Z_{2}}{2} [\nabla \vec \phi(x)]^{2} +    Z_{4}\frac{16 \pi^{2}}{4!} g \mu^{\epsilon}\!\left[\vec \phi(x)^{2} \right]^{2}\! . 
\ee
The relation between bare and renormalized quantities reads
\bea
\vec \phi_{0}(x) &=& \sqrt {Z_{2}} \,\vec \phi(x) = :  {Z_{\phi}} \,\vec \phi(x)\ , \\
m_{0}^{2} &=& \frac {Z_{1}}{Z_{2}} m^{2} =: Z_{m^{2}}  m^{2}\ ,\\
g _{0} &=& \frac{Z_{4}}{Z_{2}^{2}}  g \mu^{\epsilon} =: Z_{g}  g \mu^{\epsilon}
 .
\eea
Using perturbation theory in $g_{0}$, counter-terms are identified to render the theory UV finite. 
In dimensional regularization and minimal subtraction \cite{HooftVeltman1972}, the $Z$-factors  only depend on $g$ and $\epsilon$, and admit a Laurent series expansion of the form
\be
Z_{i} = Z_{i}(g,\epsilon) = 1+ \sum_{k=1}^{\infty}\frac{Z_{i,k}(g)}{\epsilon^{k}}\ .
\ee
Each $Z_{i,k}(g)$ is a power-series in the coupling $g$, starting at order $g^{k}$, or higher.

Three renormalization (RG) functions   can be constructed out of the three $Z$-factors. The $\beta$-function, quantifying the flow of the coupling constant, reads
\be
\beta(g) := \mu \left.\frac{\partial g}{\partial \mu}\right|_{g_{0}} = \frac{- \epsilon g}{1+ g \frac{\partial \ln (Z_{g})}{\partial g}}\ .
\ee
The RG functions associated to the anomalous dimensions are defined as  
\be
\gamma_{i}(g) := \mu   \frac {\partial }{\partial \mu} \ln( Z_{i}) = \beta(g) \frac {\partial }{\partial g} \ln( Z_{i}(g)) \ .
\ee
To leading order, the expansion of the $\beta$ function is
\be
\beta(g) = -  \epsilon g +\frac{n+8}{3}g^{2} + \ca O(g^{3})
\ .
\ee
Thus, at least for $\epsilon$ small, there is a fixed point with $\beta(g_{*}) =0$ at 
\be
g_* = \frac{3 \epsilon}{n+8} + \ca O(\epsilon^{2})\ .
\ee
It is infrared (IR) attractive, thus governs the properties of the system at large scales. This is formally deduced from the {\em correction-to-scaling} exponent $\omega$, defined as 
\be
\omega := \beta'(g_{*}) = \epsilon +  \ca O(\epsilon^{2}) \ .
\ee
The exponents $\nu$ and $\eta$ are obtained from the remaining RG functions   
\bea
\eta &=& 2 \gamma_{\phi}(g_{*}) \equiv \gamma_{2} (g_{*}) \\
\nu^{-1}  &=& 2 + \gamma_{m^{2}}(g_{*}) \equiv 2+\gamma_{1}(g_{*}) -\eta
\label{nu}
\ .
\eea
Since $g_{*} = \ca O(\epsilon)$, the perturbative expansion in $g$ is turned into a perturbative expansion in $\epsilon$. 
While the exponents $\nu$ and  $\eta$ are well-defined in the critical theory, it is not clear whether $\omega$ can be obtained from the critical theory as well. 

A different class of exponents concerns geometrical objects as the fractal dimension of lines. An example  is the self-avoiding polymer, also known as self-avoiding walk (SAW), whose radius of gyration $R_{\rm g}$ scales with its microscopic length $\ell$ as 
\be
R_{\rm g}^{\rm SAW} \sim \ell ^{\nu}\ .
\ee
Its fractal dimension is 
\be\label{df-SAW}
d_{\rm f}^{{\rm SAW}} = \frac 1\nu\ .
\ee
\begin{figure}
\Fig{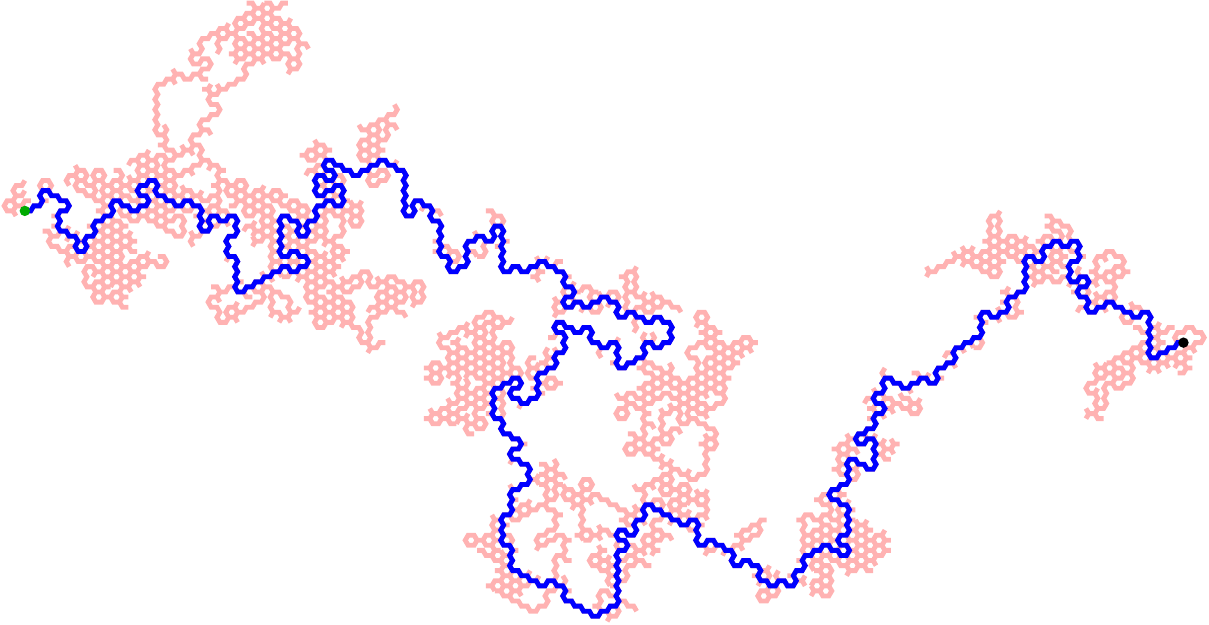}
\caption{Example of a loop-erased random walk on the hexagonal lattice with 3000 steps, starting at the black point to the right and arriving at the green point to the left.}
\label{f:LERW}
\label{nu-d-naive}
\end{figure}In general, however, $\nu$ does not yield the scaling of critical curves, but of the ensemble of all loops. This can be seen for the loop-erased random walk depicted in Fig.~\ref{f:LERW}. It is constructed   by following a random walk at time $t$, for all $t\le {\cal T}$. Whenever the walk comes back to a site it already visited, the ensuing loop is erased \cite{Lawler1980}. The remaining simple curve (blue on Fig.~\ref{f:LERW}) is the loop-erased random walk (LERW). The  trace of the underlying random walk  (RW)   is depicted in red (for the erased parts) and blue (for the non-erased part).  Its fractal dimension   is (see e.g.\ \cite{Falconer1986} Theorem 8.23)
\be
d_{\rm f}^{\rm RW} = 2 
\ee
in all dimensions $d\ge 2$, and its radius of gyration scales as  
\be
R_{\rm g}^{{\rm RW}} \sim T^{\nu}\ , \qquad \nu = \frac12\ .
\ee
The same scaling holds (by construction) for LERWs, 
\be
R_{\rm g}^{\rm {LERW}} \sim T^{\nu}\ , \qquad \nu = \frac12\ ,
\ee
but this does not tell us anything about its fractal dimension, i.e.\ the blue curve, which in $d=2$ is \cite{LawlerSchrammWerner2004}
\be
d_{\rm f} ^{\rm LERW} = \frac 5 4\ .
\ee
The latter appears in the scaling of the radius of gyration with the backbone length, i.e.\ 
\be
R_{\rm g}^{\rm {LERW}} \sim \ell^{1/d_{\rm f}}\ , 
\ee
or can be extracted by measuring the backbone length $\ell$ as a function of time, 
\be
\ell \sim T^{\phi_{\rm c}}\ ,\qquad \phi_{\rm c}=\nu d_{\rm f}\ .
\ee
While the function $\gamma_{m^{2}}$ gives us the RG-flow of the operator 
\be
{\cal E}(x):= \frac1n \sum_{i=1}^{n} \phi_{i}^{2}(x)\ , 
\ee
there is a second  $O(n)$-invariant operator bilinear in $\phi$, namely the traceless tensor operator 
\be\label{29}
\tilde {\cal E}_{ij}(x):= \phi_{i}(x) \phi_{j }(x) -  {\delta _{ij}} {\cal E}(x)\ .
\ee
By construction
\be\label{tildeEtrace}
\sum_{i} \tilde {\cal E}_{ii}(x)=0\ .
\ee 
Now consider the insertion of operators $\cal E$ and $\tilde {\cal E}$ into an expectation value. More specifically, insert  (we choose normalizations convenient for the calculations)   \be\label{26}
{\cal E}:= {\half} \int_{y}\sum _{i}\phi_{ i}^{2}(y)
 \ee
 into a diagram in perturbation theory of the form
\bea\label{e:example}
&& \left< \phi_{1}(x) \phi_{1}(z) \int _{y}\sum_i\half\phi_{i}^{2}(y)\,\rme^{{-\cal S}}\right>  =
{\parbox{1.1cm}{{\begin{tikzpicture}
\coordinate (v1) at  (0,1.25) ; \coordinate (v2) at  (0,-.25) ;  \node (x) at  (0,0)    {$\!\!\!\parbox{0mm}{$\raisebox{-3mm}[0mm][0mm]{$\scriptstyle x$}$}$};
\node (yy) at  (1,0.5)    {$\!\!\!\parbox{0mm}{$\raisebox{-1mm}[0mm][0mm]{~~$\scriptstyle y$}$}$};
\coordinate (x1) at  (0.5,0);\coordinate (y) at  (1,0.5); \coordinate (y1) at  (0.5,1) ;\node (z) at  (0,1)    {$\!\!\!\parbox{0mm}{$\raisebox{1mm}[0mm][0mm]{$\scriptstyle z$}$}$};
\fill (x) circle (1.5pt);
\fill (y) circle (1.5pt);
\fill (z) circle (1.5pt);
\draw [blue] (x) -- (x1);
\draw [blue] (y1) -- (z);
\draw [blue](0.5,0) arc (-90:90:0.5);
\end{tikzpicture}}}}  
 \nn\\
&&-\, g
{\parbox{1.6cm}{{\begin{tikzpicture}
\node (v1) at  (0,1.25){} ;
\node (v2) at  (0,-.25){} ;
\node (x) at  (0,0)    {$\!\!\!\parbox{0mm}{$\raisebox{-3mm}[0mm][0mm]{$\scriptstyle x$}$}$};
\coordinate (x1) at  (1,0) ;\coordinate (y) at  (1.5,0.5); \coordinate (y1) at  (1,1);\node (z) at  (0,1)    {$\!\!\!\parbox{0mm}{$\raisebox{1mm}[0mm][0mm]{$\scriptstyle z$}$}$};
\node (yy) at  (0.35,1)    {$\!\!\!\parbox{0mm}{$\raisebox{-2.5mm}[0mm][0mm]{~~$\scriptstyle y$}$}$};
\coordinate (y) at  (.5,1);
\fill (x) circle (1.5pt);
\fill (y) circle (1.5pt);
\fill (z) circle (1.5pt);
\fill (x1) circle (1.5pt);
\fill (y1) circle (1.5pt);
\draw [blue] (x) -- (x1);
\draw [blue] (y1) -- (z);
\draw [blue](1,0) arc (-90:90:0.5);
\draw [dashed] (x1) -- (y1);
\end{tikzpicture}}}}
~-\, g
{\parbox{1.6cm}{{\begin{tikzpicture}
\node (v1) at  (0,1.25){} ;
\node (v2) at  (0,-.25){} ;
\node (x) at  (0,0)    {$\!\!\!\parbox{0mm}{$\raisebox{-3mm}[0mm][0mm]{$\scriptstyle x$}$}$};
\coordinate (x1) at  (1,0) ;\coordinate (y) at  (1.5,0.5); \coordinate (y1) at  (1,1);\node (z) at  (0,1)    {$\!\!\!\parbox{0mm}{$\raisebox{1mm}[0mm][0mm]{$\scriptstyle z$}$}$};
\node (yy) at  (0.35,0)    {$\!\!\!\parbox{0mm}{$\raisebox{-3mm}[0mm][0mm]{~~$\scriptstyle y$}$}$};
\coordinate (y) at  (.5,0);
\fill (x) circle (1.5pt);
\fill (y) circle (1.5pt);
\fill (z) circle (1.5pt);
\fill (x1) circle (1.5pt);
\fill (y1) circle (1.5pt);
\draw [blue] (x) -- (x1);
\draw [blue] (y1) -- (z);
\draw [blue](1,0) arc (-90:90:0.5);
\draw [dashed] (x1) -- (y1);
\end{tikzpicture}}}}
~-\, g {\parbox{1.6cm}{{\begin{tikzpicture}
\node (v1) at  (0,1.25){} ;
\node (v2) at  (0,-.25){} ;
\node (x) at  (0,0)    {$\!\!\!\parbox{0mm}{$\raisebox{-3mm}[0mm][0mm]{$\scriptstyle x$}$}$};
\node (yy) at  (1.5,0.5)    {$\!\!\!\parbox{0mm}{$\raisebox{-1mm}[0mm][0mm]{~~$\scriptstyle y$}$}$};
\coordinate (y) at  (1.5,0.5);
\coordinate (x1) at  (1,0) ;\coordinate (y) at  (1.5,0.5); \coordinate (y1) at  (1,1);\node (z) at  (0,1)    {$\!\!\!\parbox{0mm}{$\raisebox{1mm}[0mm][0mm]{$\scriptstyle z$}$}$};
\fill (y) circle (1.5pt);
\fill (x) circle (1.5pt);
\fill (z) circle (1.5pt);
\fill (x1) circle (1.5pt);
\fill (y1) circle (1.5pt);
\draw [blue] (x) -- (x1);
\draw [blue] (y1) -- (z);
\draw [blue](1,0) arc (-90:90:0.5);
\draw [dashed] (x1) -- (y1);
\end{tikzpicture}}}}\nn
\\ 
&&-\,g
{\parbox{2.6cm}{{\begin{tikzpicture}
\node (v1) at  (0,1.25){} ;
\node (v2) at  (0,-.25){} ;
\node (yy) at  (0.35,1)    {$\!\!\!\parbox{0mm}{$\raisebox{-2.5mm}[0mm][0mm]{~~$\scriptstyle y$}$}$};
\coordinate (y) at  (.5,1);
\node (x) at  (0,0)    {$\!\!\!\parbox{0mm}{$\raisebox{-3mm}[0mm][0mm]{$\scriptstyle x$}$}$};
\coordinate (x1) at  (0.5,0) ;\coordinate (y1) at  (0.5,1);\coordinate (y2) at  (1.5,1) ; \node (z) at  (0,1)    {$\!\!\!\parbox{0mm}{$\raisebox{1mm}[0mm][0mm]{$\scriptstyle z$}$}$};
\coordinate (h1) at  (1,0.5) ;
\coordinate (h2) at  (1.5,0.5) ;
\fill (x) circle (1.5pt);
\fill (z) circle (1.5pt);
\fill (y) circle (1.5pt);
\fill (h1) circle (1.5pt);
\fill (h2) circle (1.5pt);
\draw [blue] (x) -- (x1);
\draw [blue] (y1) -- (z);
\draw [blue](0.5,0) arc (-90:90:0.5);
\draw [red](1.5,0.5) arc (-180:180:0.5);
\draw [dashed] (h1) -- (h2);
\end{tikzpicture}}}}~ 
-\,g
{\parbox{2.6cm}{{\begin{tikzpicture}
\node (v1) at  (0,1.25){} ;
\node (v2) at  (0,-.25){} ;
\node (yy) at  (0.35,0)    {$\!\!\!\parbox{0mm}{$\raisebox{-2.5mm}[0mm][0mm]{~~$\scriptstyle y$}$}$};
\coordinate (y) at  (.5,0);
\node (x) at  (0,0)    {$\!\!\!\parbox{0mm}{$\raisebox{-3mm}[0mm][0mm]{$\scriptstyle x$}$}$};
\coordinate (x1) at  (0.5,0) ;\coordinate (y1) at  (0.5,1);\coordinate (y2) at  (1.5,1) ; \node (z) at  (0,1)    {$\!\!\!\parbox{0mm}{$\raisebox{1mm}[0mm][0mm]{$\scriptstyle z$}$}$};
\coordinate (h1) at  (1,0.5) ;
\coordinate (h2) at  (1.5,0.5) ;
\fill (x) circle (1.5pt);
\fill (z) circle (1.5pt);
\fill (y) circle (1.5pt);
\fill (h1) circle (1.5pt);
\fill (h2) circle (1.5pt);
\draw [blue] (x) -- (x1);
\draw [blue] (y1) -- (z);
\draw [blue](0.5,0) arc (-90:90:0.5);
\draw [red](1.5,0.5) arc (-180:180:0.5);
\draw [dashed] (h1) -- (h2);
\end{tikzpicture}}}}\nn
\\
&&-\, g {\parbox{2.8cm}{{\begin{tikzpicture}
\node (v1) at  (0,1.25){} ;
\node (v2) at  (0,-.25){} ;
\node (x) at  (0,0)    {$\!\!\!\parbox{0mm}{$\raisebox{-3mm}[0mm][0mm]{$\scriptstyle x$}$}$};
\coordinate (x1) at  (0.5,0) ;
\coordinate (y) at  (2.5,0.5);
\coordinate (y1) at  (0.5,1);
\coordinate (y2) at  (1.5,1) ; 
\node (z) at  (0,1)    {$\!\!\!\parbox{0mm}{$\raisebox{1mm}[0mm][0mm]{$\scriptstyle z$}$}$};
\node (yy) at  (2.5,0.5)    {$\!\!\!\parbox{0mm}{$\raisebox{-1mm}[0mm][0mm]{~~$\scriptstyle y$}$}$};
\coordinate (h1) at  (1,0.5) ;
\coordinate (h2) at  (1.5,0.5) ;
\fill (x) circle (1.5pt);
\fill (z) circle (1.5pt);
\fill (y) circle (1.5pt);
\fill (h1) circle (1.5pt);
\fill (h2) circle (1.5pt);
\draw [blue] (x) -- (x1);
\draw [blue] (y1) -- (z);
\draw [blue](0.5,0) arc (-90:90:0.5);
\draw [red](1.5,0.5) arc (-180:180:0.5);
\draw [dashed] (h1) -- (h2);
\end{tikzpicture}}}}~~~ + \,... 
\ .
\eea
{All contributions up to 1-loop order are drawn: 
On the first line is the free-theory contribution. The insertion     of $\int_{y}\sum_i\half\phi_{i}^{2}(y)$  gives the length (in time) of the free propagator. On the second     line are   the first type of  1-loop contributions, with the insertion of $\int_{y}\sum_i\half\phi_{i}^{2}(y)$ twice in an outer line, once in a loop.  On the third and fourth line are the remaining 1-loop contributions, with the red loop counting a factor of $n$. This stems from our graphical convention to note the  $ ( \vec\phi^{2} )^{\!2}$-vertex  as 
\be
\color{black}
\big( \vec\phi^{2}\big)^{\!2} =\,   {\parbox{.6cm}{{\begin{tikzpicture}
\coordinate (v1) at  (0,0) ;
\coordinate (v2) at  (0,1) ;
\coordinate (v3) at  (.5,0) ;
\coordinate (v4) at  (.5,1) ;
\coordinate (h1) at  (0,0.5) ;
\coordinate (h2) at  (0.5,0.5) ;
\fill (h1) circle (1.5pt);
\fill (h2) circle (1.5pt);
\draw [blue] (v1) --(h1)-- (v2);
\draw [blue] (v3) --(h2)-- (v4);
\draw [dashed] (h1) -- (h2);
\end{tikzpicture}}}}~~;
\ee
contracting the two right-most lines leads to a free summation $\sum_{i}$, i.e.\ a factor of $n$ indicated in red above. 

These perturbative corrections are in one-to-one correspondence to diagrams in the high-temperature lattice expansion, where in appropriate units $g$ is set to 1. Both expansions  yield the total length of all lines, be it propagator or loop.}
\begin{figure}[t]
{\fig{1}{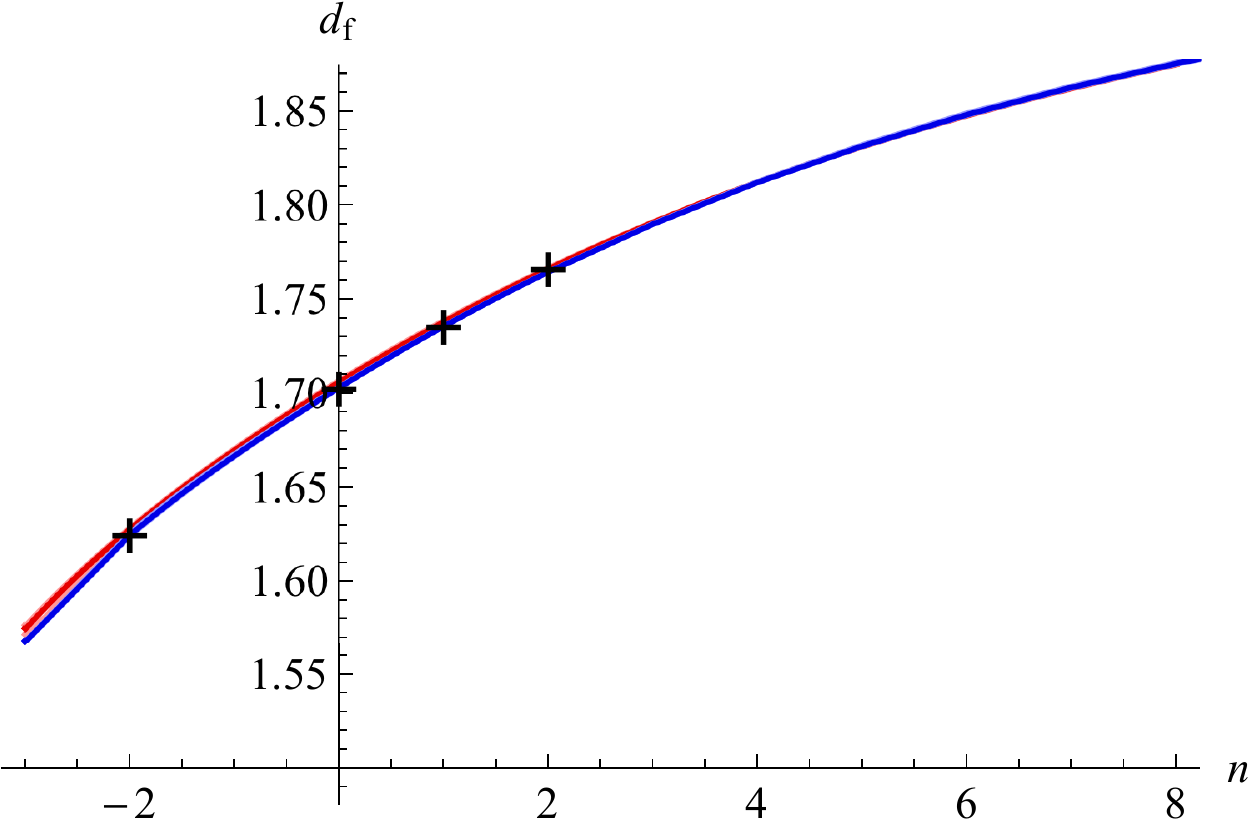}}
\smallskip

\begin{tabular}{ccccc}
\hline\hline
 ~~~~$d_{\rm f}$~~~ & ~~$n$~~ & SC & KP17 & simulation\\ \hline 
  LERW & $-2$ & $1.6243(10)$ & $1.623(6)$  & $1.62400 (5)$ \mbox{\cite{Wilson2010}}\\
 SAW & $0$ & $1.7027(10)$ &$1.7025(7)$ &$1.701847(2)$ \mbox{\cite{ClisbyDunweg2016}}\\
 Ising & $1$ & $1.7353 (10) $&$1.7352(6)$ &$1.7349 (65)$ \mbox{\cite{WinterJankeSchakel2008}}\\
 XY &$2$ & $1.7644 (10)$ & $1.7642(3)$& $1.7655(20)$ \mbox{\cite{ProkofevSvistunov2006,WinterJankeSchakel2008}}\\
 \hline \hline
\end{tabular}
\caption{Fractal dimensions of lines in dimension $d=3$. Two expansions are shown: direct (in red) and expansion for $1/d_{\rm f}$ (blue). The table compares our values to results from the literature.}
\label{tab:df-d=3}
\end{figure}

As the insertion of $\half\int_{y}\sum_{i}\phi_{i}^{2}(y)$ can be generated by deriving the action \eq{S-bare} w.r.t.\ the mass, the fractal dimension of all lines is related to $\nu$ as in \Eq{df-SAW} via
\be\label{df-tot}
d_{\rm f}^{{\rm tot}} = \frac 1\nu = 2 + \gamma_{1} (g_{*})-\eta\ .
\ee
We are now in a position to evaluate the fractal dimension of the blue line,  also termed {\em propagator line} or {\em backbone}, i.e.\ excluding  loops: This is achieved by inserting an operator proportional to $\tilde {\cal E}_{ij}$. To be specific, we consider the insertion of  
\be\label{34}
\tilde {\cal E} := \frac 12 \int_{y}    \phi_{1}^{2}(y)-  \phi_{2}^{2}(y) \ .
\ee
 This is, with a normalization convenient for our calculations,   the integrated form of $\tilde {\cal E}_{11}-\tilde {\cal E}_{22}$ defined in \Eq{29}.
When evaluated in a line with  index ``1'' (the correlation function of $\langle\phi_1 \phi_1\rangle$), i.e.\ in the  blue  line in \Eq{e:example} which is connected to the two external points, the result is the same as for the insertion \eq{26}. 
On the other hand, when inserted into a loop (drawn in red), where the sum over indices is unrestricted, it vanishes.

{Let us give some background information:
In the  $O(n)$-model,  the number of components $n$ is a priori a positive integer, but can analytically be continued to arbitrary $n$. Two non-positive values of $n$ merit special attention: 
$n=0$ corresponds to self-avoiding polymers, as shown by De Gennes \cite{DeGennes1972}. Here the propagator line (in blue) is interpreted as the self-avoiding polymer, and the red loops are absent. 
Focusing on lattice configurations with one self-intersection, see \Eq{e:example}, the choice of $g=1$ cancels the free-theory result, giving total weight 0 for self-intersecting paths -- as expected.  
The second case of interest is $n=-2$, and corresponds to loop-erased random walks \cite{WieseFedorenko2018,WieseFedorenko2019}. Here all perturbative terms  $\sim g$ cancel, as the propagator of a loop-erased random walk is identical to that of a random walk. To our advantage, we can  equivalently use the cancelation of 
the first two lines (as for self-avoiding polymers). Then  the random walk is redrawn in a way making visible the loop-erased random walk (in blue) and the erased loop (in red), allowing to extract the fractal dimension of the loop-erased random walk via the operator $\tilde {\cal E}$ as given in \Eq{34}. For details, we refer to Refs.~\cite{WieseFedorenko2018,WieseFedorenko2019}.} 

The operator $\tilde {\cal E}$ can be renormalized multiplicatively, by considering the insertion 
\be
\delta S =  \lambda  \frac {Z_{\tilde {\cal E}}}2 \int_{y}   \phi_{0,1}^{2}(y)-  \phi_{0,2}^{2}(y) \ ,
\ee
where $\phi_{0,i}$ denotes the $i$-th component of the bare field $\phi_0$.
As a result,  the fractal dimension of the propagator (or backbone) line is given by 
\bea\label{df}
d_{\rm f}  &=& 2 + \gamma_{\tilde {\cal E}} (g_{*})-\eta\ ,\\
\gamma_{\tilde {\cal E}} &:=&  \mu   \frac {\partial }{\partial \mu} \ln( Z_{\tilde {\cal E}}) = \beta(g) \frac {\partial }{\partial g} \ln( Z_{\tilde {\cal E}}(g)) \ .
\eea\begin{figure}[t]
{\fig{1.00}{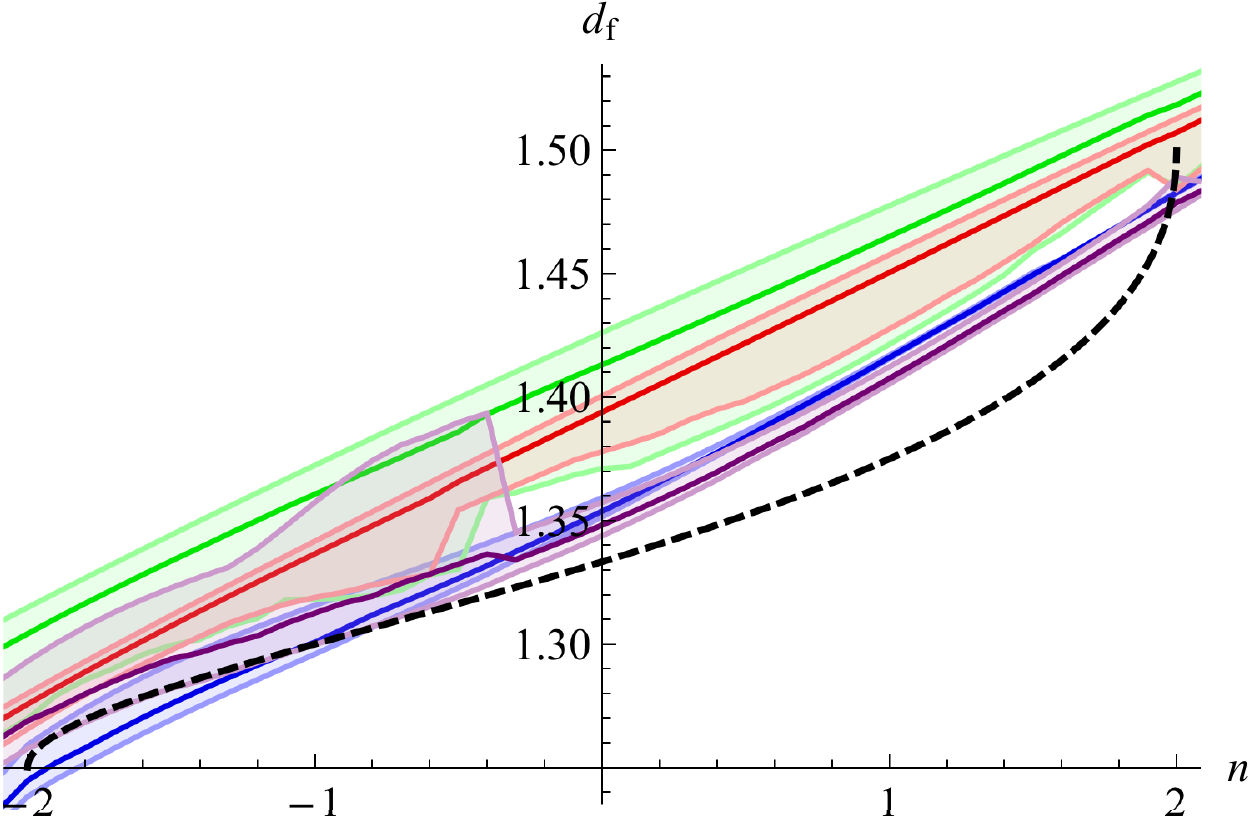}}
\smallskip

\begin{tabular}{ccccc}
\hline
\hline
 ~~~~$d_{\rm f}$~~~ & ~~$n$~~ & SC & KP17 & CFT\\ \hline  LERW & $-2$ & $1.244(6)$ & $1.188(55)$  &$5/4=1.25$\\
 SAW & $0$ & $ 1.354(5)$ &$1.350(8)$ &$4/3\simeq 1.333$\\
 Ising & $1$ & $1.416(1)$ &$1.413(7)$ &$11/8 = 1.375$\\
 XY &$2$ & $1.482(1)$ & $1.480(4)$& $3/2=1.5$\\
 \hline 
 \hline
\end{tabular}
\caption{The fractal dimension of lines in dimension $d=2$, as extracted from field theory (colored), and compared to exact results (black dashed line). The different curves are from resummation of $d_{\rm f}$ (blue), $d_{\rm f}^{-1}$ (red), $d_{\rm f}^2$ (cyan), and   $d_{\rm f}^{-2}$ (green). The table compares   the result of our different schemes, with the direct expansion of $d_{\rm f}$ used for SC. Note that the error given is the error of the expansion in one scheme. Comparing different expansion schemes, we estimate the overall error to be of order $0.05$.}
\label{f:df:d=2}
\end{figure}The explicit result to 6-loop order is given below in \Eq{gamma-tilde-E}.
In the literature \cite{Kirkham1981,Amit,Gracey2002,ShimadaHikami2016,WieseFedorenko2018} one also finds the ratio
\be\label{38}
\phi_{\rm c}(n):= \nu d_{\rm f} \equiv \frac{d_{\rm f}}{d_{\rm f}^{\rm tot}}\equiv \frac{2 + \gamma_{\tilde {\cal E}} (g_{*})-\eta}{2 + \gamma_{1} (g_{*})-\eta}\ .
\ee
It is known as {\em crossover exponent}, since it describes the crossover from a broken symmetry $O(k)$, $k<n$, to $O(n)$. We will review this in section \ref{s:crossover} below. 
Since for $n=0$ all loops are absent, the two fractal dimensions coincide. For positive $n$, the fractal formed by backbone plus loops is larger than the backbone, and we expect $d_{\rm f}^{\rm tot}>d_{\rm f}$. Translated to $\phi_{\rm c}(n)$ this implies
\be
\phi_{\rm c}(0) = 1\ , \qquad \phi_{\rm c}'(n)>0\ .
\ee
The last relation, which is stronger than $d_{\rm f}^{\rm tot}>d_{\rm f}$ is expected since the derivative w.r.t.\ $n$ counts loops which are added to the fractal when increasing $n$, which should be positive. 

Let us now turn to a comparison of the fractal dimension  given by \Eq{df} with numerical simulations. 
There are four systems for which simulations are available (summarized in Fig.~\ref{tab:df-d=3}). \begin{enumerate}
\item [(i)]loop-erased random walks: As shown in \cite{WieseFedorenko2018} this is given by $n=-2$, in all dimensions. 
\item [(ii)] self-avoiding polymers: $n=0$. Here $d_{\rm f}\equiv 1/\nu$. 
\item [(iii)] Ising model: $n=1$.
\item [(iv)] XY-model: $n=2$.
\end{enumerate}
Simulations for the Ising and XY model are  performed on the lattice \cite{ProkofevSvistunov2006,WinterJankeSchakel2008}, by considering the high-temperature expansion which allows the authors to distinguish between propagator lines and loops, similar to our discussion of the perturbative expansion \eq{e:example}.

In all cases, the agreement of our RG results with simulations in $d=3$ is excellent, firmly establishing that     the  appropriate operator was identified. In dimension $d=2$ (shown on Fig.~\ref{f:df:d=2}), different resummation procedures (see below) yield different results, showing that extrapolations down to $d=2$ are difficult. This can be understood from the non-analytic behavior of the exact result close to $n=\pm 2$. It is even more pronounced for the exponent $\nu$ (see figure \ref{f:nu:d=2} below), which diverges with a square-root singularity at $n=2$. We will come back to this issue in section \ref{s:The limit of d=2 checked against conformal field theory}.

The remainder of this article is organized as follows: In section \ref{s:The RG function gamma-t-E} we give the explicit result for the new RG-function  $\gamma_{\tilde {\cal E}}$. Section \ref{s:A self-consistent resummation procedure}  introduces a self-consistent resummation procedure as a (fast) alternative to the elaborate scheme of Ref.~\cite{KompanietsPanzer2017}. 
In the next two sections we   discuss   in more detail the dimension of curves, and their relation to the  {\em crossover exponent} (section  \ref{s:crossover}) and loop-erased random walks (section \ref{s:Loop-erased random walks}). Section \ref{s:The limit of d=2 checked against conformal field theory} tests the $\epsilon$-expansion against analytic results in dimension $d=2$, allowing us to identify the most suitable variables for the resummation procedure.  This allows us to give in   section \ref{s:estimates in d=3}   improved predictions for all relevant exponents in dimension $d=3$.  
Section \ref{s:Connectiontolarge-Nexpansion} makes the connection to known results from the large-$n$ expansion, which serves as a non-trivial test of our results. We conclude in section \ref{s:Conclusion}.

\begin{widetext}
\section{The RG function $\gamma_{\tilde {\cal E}}$}
\label{s:The RG function gamma-t-E}
The RG function  $\gamma_{\tilde {\cal E}}$ 
to 6-loop order, evaluated at the fixed point,  reads 
(with $d=4-\epsilon$) 
\bea
\label{gamma-tilde-E}\gamma_{\tilde {\cal E}}&=&
-\frac{2 \epsilon }{n+8}
+\epsilon ^2\left[\frac{ \left(n^2-4 n-36\right)}{(n+8)^3}\right]
+\epsilon ^3 \left[\frac{24 (5 n+22) \zeta_3}{(n+8)^4}+\frac{n^4+45 n^3+190 n^2-144 n-1568}{2 (n+8)^5}\right]\nn\\
&+&\epsilon ^4 \Bigg[-\frac{80 \left(2 n^2+55 n+186\right) \zeta_5}{(n+8)^5}+\frac{18 (5 n+22) \zeta_4}{(n+8)^4}-\frac{\left(n^5+16 n^4+808 n^3+3624 n^2-6240 n-30528\right) \zeta_3}{2 (n+8)^6}
\w
+\frac{2 n^6+135 n^5+3672 n^4+26568 n^3+87528 n^2+123264 n+6016}{8 (n+8)^7}\Bigg]\nn
\\
&+&\epsilon ^5 \Bigg[\frac{882 \left(14 n^2+189 n+526\right) \zeta_7}{(n+8)^6}-\frac{100 \left(2 n^2+55 n+186\right) \zeta_6}{(n+8)^5}-\frac{4 \left(5 n^4+6 n^3+3444 n^2+26824 n+46752\right) \zeta_3^2}{(n+8)^7}\nn
\eea
\bea
&&+\frac{\left(895 n^4+20194 n^3+73636 n^2-68712 n-403392\right) \zeta_5}{(n+8)^7}-\frac{3 \left(n^5+16 n^4+808 n^3+3624 n^2-6240 n-30528\right) \zeta_4}{8 (n+8)^6}
\w
+\frac{\left(n^7-36 n^6-176 n^5-35336 n^4-336080 n^3-842848 n^2+394624 n+2870528\right) \zeta_3}{4 (n+8)^8}
\w
+\frac{4 n^8+367 n^7+13724 n^6+275384 n^5+2162776 n^4+9337408 n^3+25225728 n^2+38978560 n+22308864}{32 (n+8)^9}\Bigg]\nn\\
&+&\epsilon ^6 \Bigg[
-\frac{64 \left(1819 n^3+97823 n^2+901051 n+2150774\right) \zeta_9}{9 (n+8)^7}
-\frac{512 \left(n^3+65 n^2+619 n+1502\right) \zeta_3^3}{(n+8)^7}
\w
-\frac{216 \left(42 n^3+2279 n^2+21282 n+50512\right) \zeta_{3,5}}{5 (n+8)^7}
+\frac{9 \left(28882 n^3+780579 n^2+5963882 n+13076112\right) \zeta_8}{10 (n+8)^7}
\w
-\frac{24 \left(59 n^4+5320 n^3+62044 n^2+364256 n+790368\right) \zeta_3 \zeta_5}{(n+8)^8}
\w
-\frac{\left(3679 n^5+605258 n^4+8044820 n^3+25012072 n^2-16957632 n-109427520\right) \zeta_7}{8 (n+8)^8}
\w
-\frac{6 \left(5 n^4+6 n^3+3444 n^2+26824 n+46752\right) \zeta_3 \zeta_4}{(n+8)^7}
+\frac{5 \left(865 n^4+19342 n^3+64708 n^2-109416 n-470976\right) \zeta_6}{4 (n+8)^7}
\w
+\frac{\left(553 n^6+9206 n^5+193932 n^4+341288 n^3-11260928 n^2-64278912 n-99677184\right) \zeta_3^2}{2 (n+8)^9}
\w
+\frac{\left(-3 n^8-104 n^7-13210 n^6-100464 n^5+2802392 n^4+27327488 n^3+78105408 n^2+46518912 n-78244864\right) \zeta_5}{8 (n+8)^9}
\w
+\frac{3 \left(n^7-36 n^6-176 n^5-35336 n^4-336080 n^3-842848 n^2+394624 n+2870528\right) \zeta_4}{16 (n+8)^8}
\w
+\Big(n^9+100 n^8+979 n^7+54758 n^6-770188 n^5-15180440 n^4-80189984 n^3-169245120 n^2
\w 
\qquad \qquad \qquad -68332544 n+162652160\Big)\frac{ \zeta_3}{8 (n+8)^{10}}
\w
+\Big(8 n^{10}+927 n^9+48746 n^8+1370920 n^7+22319040 n^6+172596192 n^5+774280256 n^4+2372987392 n^3
\w
\qquad \qquad \qquad +5281970176 n^2+7489404928 n+4525309952\Big)\frac{1}{128 (n+8)^{11}}\Bigg] + \ca O (\epsilon ^7 )
\eea

\end{widetext}
This agrees with Kirkham \cite{Kirkham1981} Eq.~(12) up to 4-loop order.
The constant $\zeta_{3,5}$ is defined as 
\be
\zeta_{3,5}:= \sum_{1\le n< m} \frac1{n^{3}m^5} \approx 0.037707673\ .
\ee
For $n=-2$ to $2$,  numerical values  of $\gamma_{\tilde {\cal E}}$ and $d_{\rm f}$ are given in 
table \ref{tab-gamma-E}.

\begin{table*}
	\centering	\caption{Numerical values for the $6$-loop RG function $\gamma_{\tilde {\cal E}}(\epsilon)$ at the fixed point.}
		\label{tab-gamma-E}
	\begin{tabular}{cl}
	\toprule
		$\quad n \quad$ & $\qquad\qquad\qquad\qquad\qquad\qquad\qquad\gamma_{\tilde {\cal E}}(\epsilon)$   \\
	\midrule
		$-2$ & $-0.333333 \epsilon - 0.111111 \epsilon^2 + 0.211568 \epsilon^3 - 
 0.611186 \epsilon^4 + 2.43354 \epsilon^5 - 11.7939 \epsilon^6+\ca O(\epsilon^7)$ \\
		$-1$ & $-0.285714 \epsilon - 0.090379 \epsilon^2 + 0.166245 \epsilon^3 - 
 0.416899 \epsilon^4 + 1.50701 \epsilon^5 - 6.60415 \epsilon^6+\ca O(\epsilon^7)$ \\
		$0$ & $-0.25  \epsilon - 0.0703125 \epsilon^2 + 0.131027 \epsilon^3 - 
 0.29588 \epsilon^4 + 0.982638 \epsilon^5 - 3.94648 \epsilon^6+\ca O(\epsilon^7)$ \\
		$1$ & $-0.222222 \epsilon - 0.0534979 \epsilon^2 + 0.106224 \epsilon^3 - 
 0.218192 \epsilon^4 + 0.673348 \epsilon^5 - 2.50444 \epsilon^6+\ca O(\epsilon^7)$ \\
		$2$ & $-0.2  \epsilon - 0.04  \epsilon^2 + 0.088718 \epsilon^3 - 
 0.165781 \epsilon^4 + 0.481055 \epsilon^5 - 1.67071 \epsilon^6+\ca O(\epsilon^7)$ \\
	\bottomrule
	\end{tabular}\end{table*}\begin{table*}
	\centering	\caption{Numerical values for the $6$-loop fractal dimension $d_{\rm f}(\epsilon)$. }
		\label{tab-df}
	\begin{tabular}{cl}
	\toprule
		$\quad n \quad$ & $\qquad\qquad\qquad\qquad\qquad\qquad\qquad d_{\rm f}(\epsilon)$   \\
	\midrule
$-2$ & 
$2-0.333333 \epsilon - 0.111111 \epsilon^2 + 0.211568 \epsilon^3 -  0.611186 \epsilon^4 + 2.43354 \epsilon^5 - 11.7939 \epsilon^6+\ca O(\epsilon^7)$\\
$-1$ & 
$2-0.285714 \epsilon - 0.100583 \epsilon^2 + 0.155051 \epsilon^3 -  0.410163 \epsilon^4 + 1.48492 \epsilon^5 - 6.52249 \epsilon^6+\ca O(\epsilon^7)$\\
$0$ & 
$2-0.25 \epsilon - 0.0859375 \epsilon^2 + 0.114425 \epsilon^3 -  0.287513 \epsilon^4 + 0.956133 \epsilon^5 - 3.85575 \epsilon^6+\ca O(\epsilon^7)$\\
$1$ & 
$2-0.222222 \epsilon - 0.0720165 \epsilon^2 + 0.0875336 \epsilon^3 -  0.209864 \epsilon^4 + 0.647691 \epsilon^5 - 2.42316 \epsilon^6+\ca O(\epsilon^7)$\\
$2$ & 
$2-0.2 \epsilon - 0.06 \epsilon^2 + 0.069718 \epsilon^3 -  0.157887 \epsilon^4 + 0.457846 \epsilon^5 - 1.60209 \epsilon^6+\ca O(\epsilon^7)$\\
 	\bottomrule
	\end{tabular}\end{table*}
\section{A self-consistent resummation procedure}
\label{s:A self-consistent resummation procedure}

\begin{figure}[b]
\fig{1}{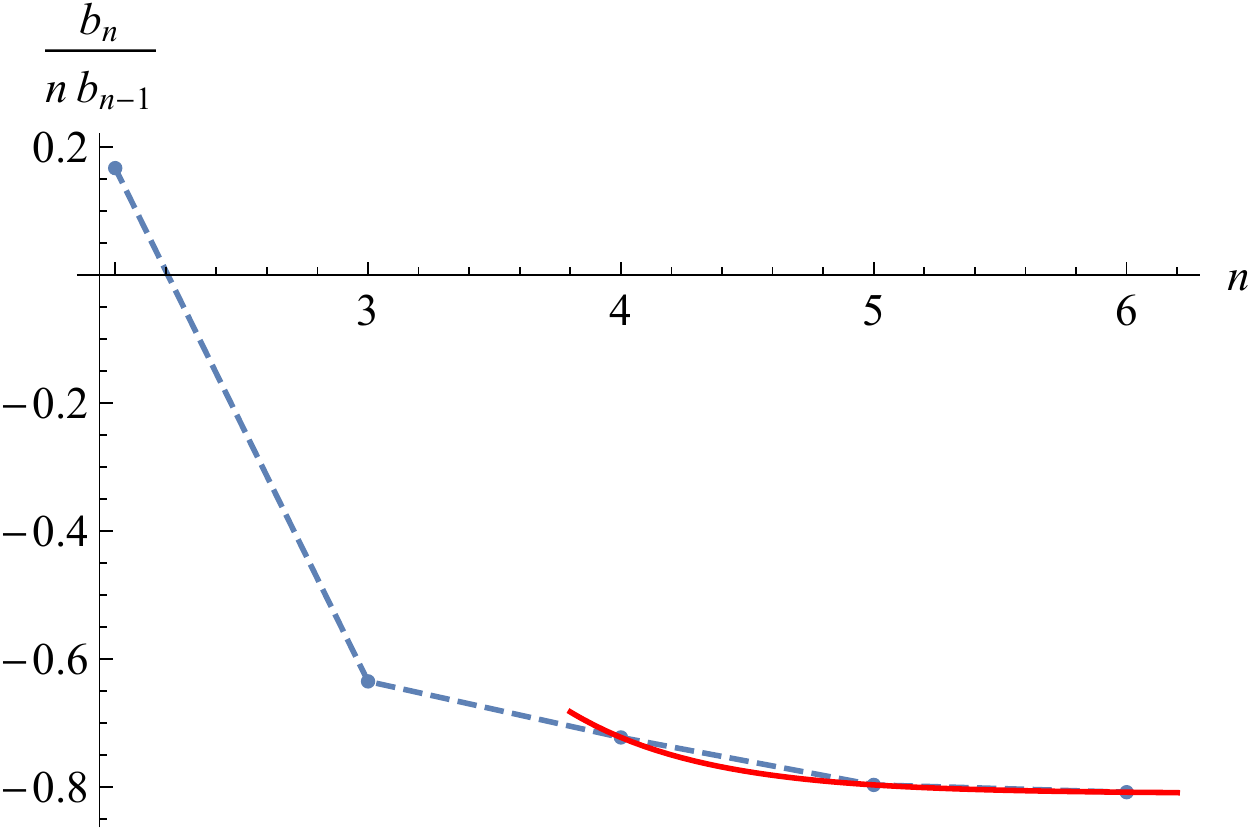}
\caption{The ratios $r_{n}$ as given in Eq.~(\ref{rn}) for $\alpha=0$, and the fit to \Eq{fit}.}
\label{f:bnratios}
\end{figure}
\begin{figure}[b]
\setlength{\unitlength}{0.05\textwidth}
\begin{picture}(10,6)
\put(0,0){\fig{1.00}{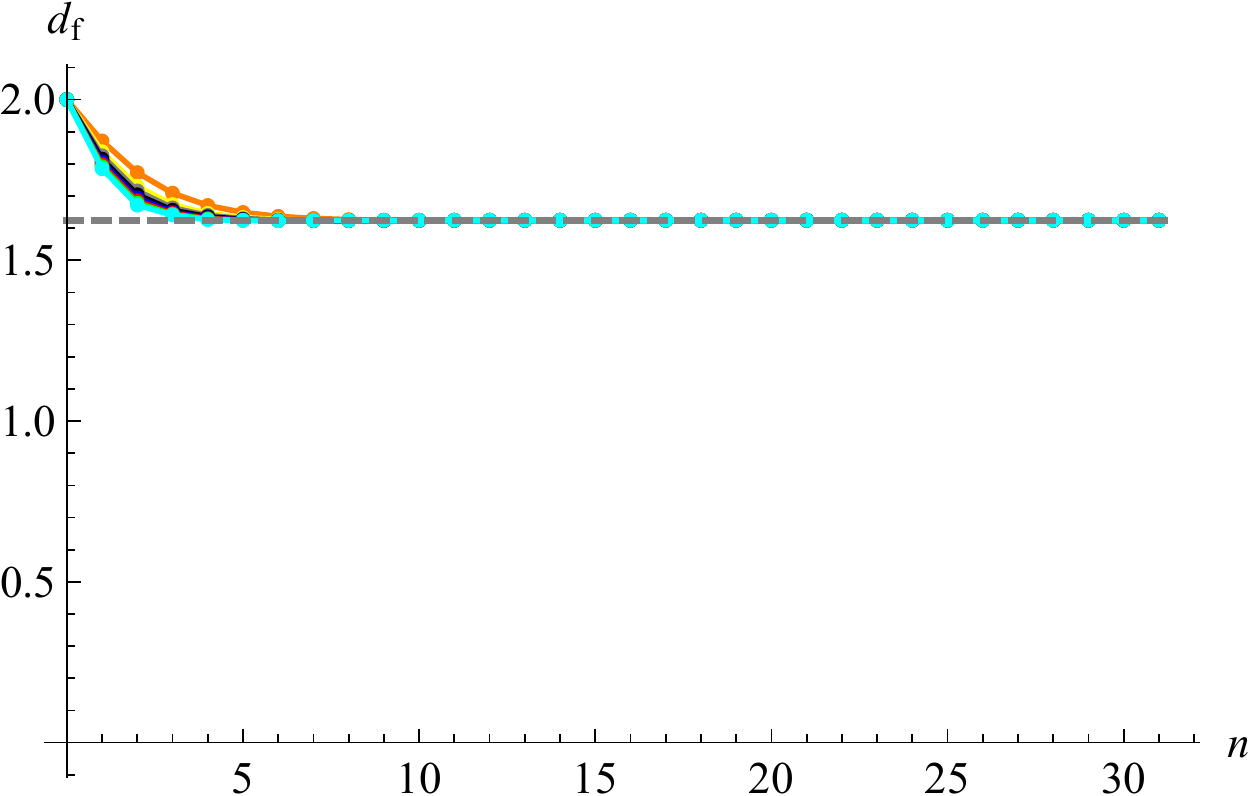}}
\put(2.1,0.5){\fig{0.7}{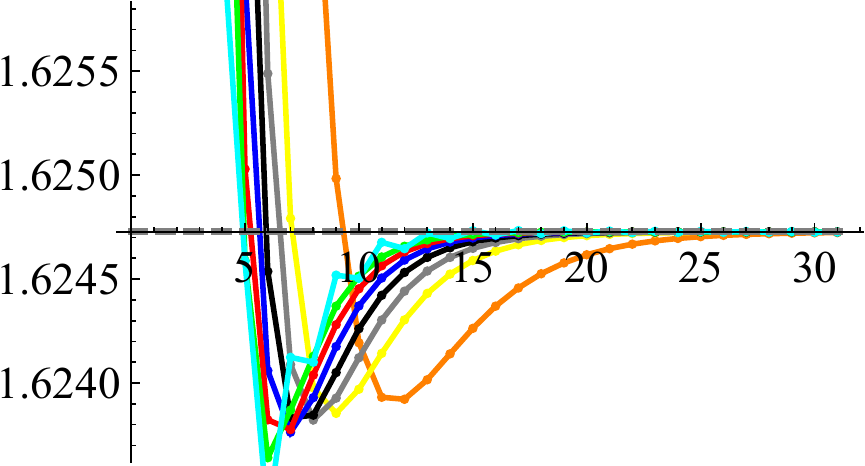}}
\end{picture}\caption{Resummation of $d_{\rm f}$ for LERWs ($n=-2$) in  $d=3$, as a function of the series-order $n$, setting $\alpha=0$. One sees that the resummed series converges, for all assumed values of the branch cut, with orange $z_{\rm bc}=0.3/a$ to green with $z_{{\rm bc}}=1/a$, ending with cyan $z_{{\rm bc}}=1.1 /a$, which clearly sits inside the supposed branch cut, which oscillates, and for which only the real part is shown.}
\label{f:dfofn}
\end{figure}

\begin{figure}\fig{1.00}{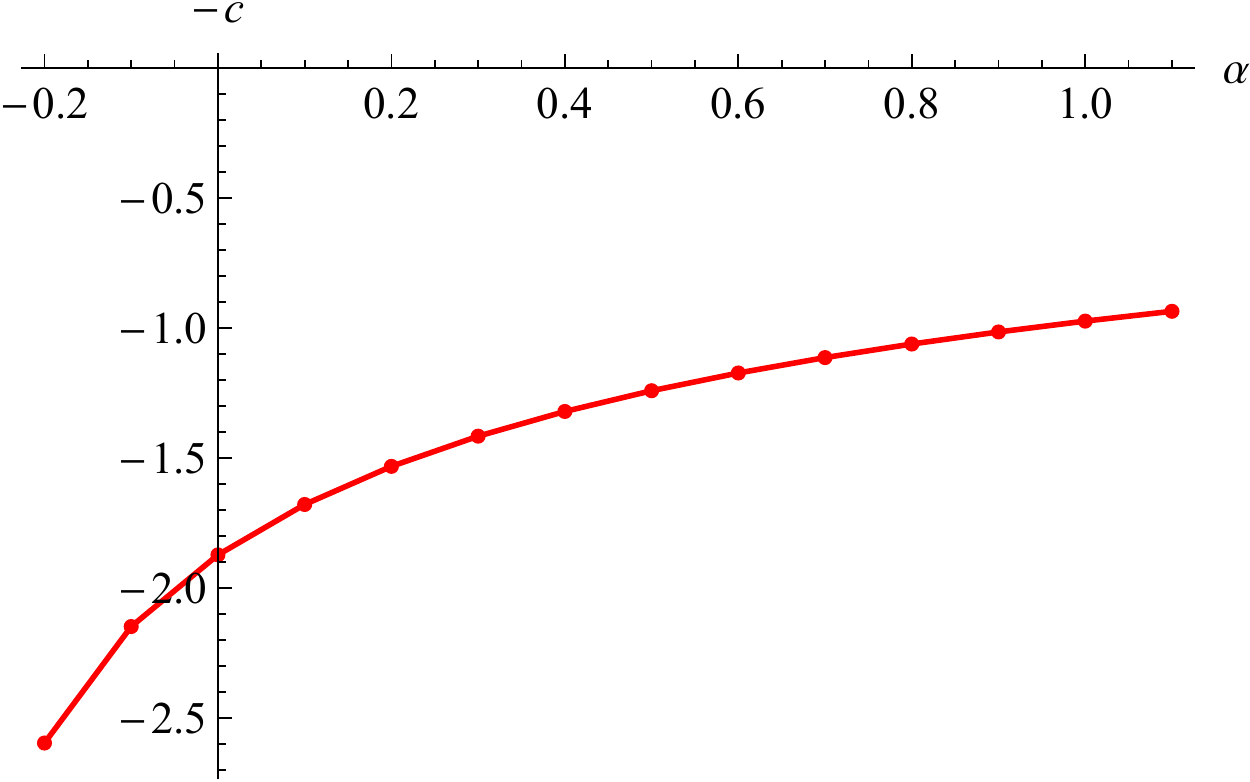}
\caption{Minus the exponential decay constant $c$ from Eq.~\eq{fit}.}
\label{f:conv-c}
\end{figure}

There are many resummation procedures \cite{GuidaZinn-Justin1998,Kleinert2000};  we show results based on the Borel-resummation method proposed in Ref.~\cite{KompanietsPanzer2017}  and denoted   KP17. We also propose a different approach, using a self-consistent (SC) resummation:
Consider an exponent or observable $\kappa(\epsilon)$, with series expansion
\be
\kappa(\epsilon) = \sum_{n=0}^{\infty} b_{n} \epsilon^{n}\ .
\ee
Suppose that $b_{n}$ has the asymptotic form 
\be
b_{n} = c_{0}   a ^{n} n! n^{\alpha}\ .
\ee
Then 
\be\label{rn}
r_{n}:=\frac{b_{n}}{b_{n-1}}\frac1n \left(\frac n{n-1}\right)^{\alpha} =  a + \delta a(n)
\ .
\ee
Further suppose that, with $c>0$
\be\label{fit}
\delta a(n) = b \,\rme^{-c n}\ .
\ee
This ansatz can be used to fit the last three elements of the table of $r_{n}$ (at 6-loop order this is $r_{2}, ... ,r_{6}$) to the three parameters $a$, $b$, and $c$. The value of $a$ is our best estimate for 
the inverse of the  branch-cut location in the inverse Borel transform. Having established a fit allows us to estimate the ratios $r_{i}$ with $i$ larger than the order to which we calculated. It in turn fixes $b_{n}$ to the same   order, in practice up to order $28...40$ using double precision, and depending on the series. An example studying the fractal dimension of LERWs is given on Figs.~\ref{f:bnratios} and \ref{f:dfofn}, for $\alpha=0$. In general, the fit \eq{fit} is possible only for a certain range of $\alpha$. The fit fails if the three chosen ratios $r_n$ are not monotone, as the exponential function then grows. As a consequence, in this case the SC scheme makes no prediction, and  we leave the corresponding table entries empty.  Different fitting forms could be proposed and tested, e.g.\ to account for such a non-monotone behavior. We restricted our tests to an algebraic decay, but  no benefit could be extracted from the latter. We believe that  the advantage of the ansatz  \eq{rn} is its fast convergence, which is lost for an ansatz with algebraic decay.

We can still use our freedom to choose $\alpha$, which also leads to different values of the exponential decay $c$ given in Fig.~\ref{f:conv-c}. Our approach is to try with all values of $\alpha$ for which a fit of the form \eq{fit} is feasible. The result is shown on Fig.~\ref{f:z-alpha}: Apart from error bars {\em of the procedure}, we obtain the mid-range and the mean of all obtained exponents as the {\em centered} and {\em best} estimates. Note that when the allowed range of $\alpha$ is small, the estimated error bars are also small, since the estimate varies continuously with $\alpha$. Thus a small  error bar may indicate a robust series and indeed a small error, or a series which is delicate to resum. As a consequence,  error bars of this method have to be taken with a grain of salt.  The method of KP17 \cite{KompanietsPanzer2017} does not suffer from this artifact.

\section{Dimension of curves and Crossover Exponent}
\label{s:crossover}
\begin{table}[t]
\centering\caption{Numerical values for the $6$-loop $\phi'_c(0)$ .}	\label{tab:dphic-numeric}	\begin{tabular}{cccc}
	\toprule
		$\quad d \quad$ & SC & KP17  & exact \\
	\midrule
$0$ & $0.70(18)$ &   $-$ \\
$1$ & $0.44(6)$ & $\quad 0.58(12)$\\
$2$ & $0.239(10)$& $\quad 0.262(10)$ & ~~$ \displaystyle \frac{3}{4 \pi } \simeq 0.238732$\\
$3$ & $0.0912(7)$& $\quad 0.0925(4)$\\
$4$ & 0 & 0 & 0 \\
 	\bottomrule
	\end{tabular}\end{table}
Following the classic book by Amit \cite{Amit}, (for more references see \cite{Kirkham1981,KiskisNarayananVranas1993,ShimadaHikami2016})
the crossover exponent arises for the following question: Consider the anisotropic $O(n)$ model, where the first $k<n$ components have a mass $m_{1}^{2}$, and the remaining $n-k$ components have a mass $m_{2}^{2}$ (we suppressed the index 0 for the bare objects for convenience of notation)
\bea\label{48}
 {{\cal S}}   &=&  \int_x \,  \frac{m^{2}_1}2 \sum_{i=1}^{k} \phi_{i}(x)^{2}+ \frac{m^{2}_2}2 \sum_{i=k+1}^{n} \phi_{i}(x)^{2} +\frac1{2} [\nabla \vec \phi(x)]^{2}\nn\\
&& \quad +     \frac{16\pi^{2}}{4!} g\!\left[\vec \phi(x)^{2} \right]^{2}\ . 
\eea
This form arises in mean-field theory, when coarse graining a $n$-component model with anisotropy. Consider   $m_{1}^{2}< m_{2}^{2}$, i.e.\ $\lambda := m_2^2-m_1^2>0$. The corresponding phase diagram is shown on figure \ref{f:crossover-phase-diagram}.  When lowering the temperature,  the $k$ first modes will become massless before the remaining ones, and one arrives at an effective $O(k)$ model. In the opposite case,  $m_{1}^{2}> m_{2}^{2}$, the remaining $n-k$ modes become massless first, resulting in a critical $O(n-k)$ model, while for $m_{1}^{2}=m_{2}^{2}$ all modes becomes massless at the same temperature. 

Let us rewrite the quadratic (derivative free) terms in \Eq{48}  as 
\be
  {{\cal S}_{m^{2}}}   = \frac{m^{2}}2  \vec \phi(x)^{2} -\frac \lambda 2 \tilde {\cal E}\ ,
\ee
where
\bea
m^{2} &:=& \frac{ k m_{1}^{2} +(n-k) m_{2}^{2}}{n}\ , \\
\lambda &:=& m_{2}^{2} - m_{1}^{2}\ , \\
 \tilde {\cal E} &=& \frac{1}n\left[ (n-k) \sum_{i=1}^{k} \phi_{i}(x)^{2} -k  \sum_{i=k+1}^{n} \phi_{i}(x)^{2}\right]\ .\qquad \label{tildeE-alt}
\eea
Further denote the distance to the critical point by
\be
t:= \frac {T-T_{{\rm c},n}}{T_{{\rm c},n}}\ .
\ee
Then any thermodynamic observable, as e.g.\ the longitudinal susceptibility, will assume a scaling form  with $t$ as
\be
\chi_{L}^{-1}(t,g) = t^{\gamma} f\left(\frac{\lambda}{t^{\phi_{\rm c}}} \right)\ .
\ee
The function $f$ is the crossover 
function, while $\phi_{\rm c}$ is  
the {\em crossover exponent}. It is the ratio of dimensions between $\lambda$ and $m^{2}$, namely
\be\label{54}
\phi_{\rm c} = \frac{\mbox{dim}_{\mu} (\lambda ) }{\mbox{dim}_{\mu} (m^{2} )} = \frac{2 + \gamma_{\tilde {\cal E}} (g_{*})-\eta}{2 + \gamma_{1} (g_{*})-\eta}\ .
\ee
In the numerator is the renormalization of $\tilde {\cal E}$ as given by \Eq{tildeE-alt}, and which sits in the same representation as $\tilde {\cal E}_{{i,j}}$ defined in \Eq{29} or $\tilde {\cal E}$  defined in \Eq{34} (thus the same notation for all these objects), and which is the fractal dimen- \begin{widetext}\noindent sion $d_{{\rm f}}$ of the backbone, as given in \Eq{df} . The denominator is $\nu^{-1}=d_{\rm f}^{\rm tot}$, as introduced in \Eq{nu}. This allows us to rewrite $\phi_{c}$ as in \Eq{38} as  
\be
\phi_{\rm c} = \frac{d_{\rm f}}{d_{\rm f}^{\rm tot}}=\nu d_{\rm f}\ .
\ee
\enlargethispage{5cm}
Its series expansion reads 
\bea
\phi_{\rm c} &=&
1 + \frac{\epsilon  n}{2 (n+8)}
+\frac{\epsilon ^2 n \left(n^2+24 n+68\right)}{4 (n+8)^3}
+\epsilon ^3 \bigg[
-\frac{6 n (5 n+22) \zeta_3}{(n+8)^4}
+\frac{n \left(n^4+48 n^3+788 n^2+3472 n+5024\right)}{8 (n+8)^5}\bigg]\nn\\
&+&\epsilon ^4 \bigg[
\frac{20 n \left(2 n^2+55 n+186\right) \zeta_5}{(n+8)^5}
-\frac{9 n (5 n+22) \zeta_4}{2 (n+8)^4}
-\frac{n \left(n^4-13 n^3+544 n^2+4716 n+8360\right) \zeta_3}{(n+8)^6}
\nn\\
&&+\frac{n \left(n^6+72 n^5+2085 n^4+28412 n^3+147108 n^2+337152 n+306240\right)}{16 (n+8)^7}
\bigg]\nn
\eea
\enlargethispage{5cm}
\bea
&+&\epsilon ^5 \bigg[
-\frac{441 n \left(14 n^2+189 n+526\right) \zeta_7}{2 (n+8)^6}
+\frac{25 n \left(2 n^2+55 n+186\right) \zeta_6}{(n+8)^5}
+\frac{2 n \left(4 n^4+39 n^3+2028 n^2+14468 n+24528\right) \zeta_3^2}{(n+8)^7}
\nn\\
&&+\frac{n \left(-230 n^4-2857 n^3+33832 n^2+280596 n+466016\right) \zeta_5}{2 (n+8)^7}
-\frac{3 n \left(n^4-13 n^3+544 n^2+4716 n+8360\right) \zeta_4}{4 (n+8)^6}
\w
-\frac{n \left(9 n^5-661 n^4+7584 n^3+125232 n^2+465592 n+554064\right) \zeta_3}{(n+8)^8}
\w
+\frac{n \left(n^8+96 n^7+4154 n^6+95668 n^5+1177480 n^4+6723904 n^3+19390624 n^2+28388096 n+17677824\right)}{32 (n+8)^9}
\bigg]\nn\\
&+&\epsilon ^6 \bigg[
\frac{16 n \left(1819 n^3+97823 n^2+901051 n+2150774\right) \zeta_9}{9 (n+8)^7}
+\frac{128 n \left(n^3+65 n^2+619 n+1502\right) \zeta_3^3}{(n+8)^7}
\w
-\frac{9 n \left(28882 n^3+780579 n^2+5963882 n+13076112\right) \zeta_8}{40 (n+8)^7}
+\frac{54 n \left(42 n^3+2279 n^2+21282 n+50512\right) \zeta_{3,5}}{5 (n+8)^7}
\w
+\frac{12 n \left(13 n^4+2288 n^3+28088 n^2+172816 n+385584\right) \zeta_3 \zeta_5}{(n+8)^8}
\w
+\frac{n \left(1136 n^5+174529 n^4+1284304 n^3-8699596 n^2-73803936 n-120419232\right) \zeta_7}{16 (n+8)^8}
\w
{ +}\frac{3 n \left(4 n^4+39 n^3+2028 n^2+14468 n+24528\right) \zeta_3 \zeta_4}{(n+8)^7}
\w
-\frac{5 n \left(215 n^4+2431 n^3-38296 n^2-300948 n-499808\right) \zeta_6}{8 (n+8)^7}
\w
+\frac{n \left(-140 n^6-471 n^5-2192 n^4+947100 n^3+11661984 n^2+46428608 n+61839872\right) \zeta_3^2}{4 (n+8)^9}
\w
+\frac{n \left(-6 n^7+348 n^6-30199 n^5-656384 n^4-615916 n^3+21367744 n^2+87069536 n+100818688\right) \zeta_5}{8 (n+8)^9}
\w
-\frac{3 n \left(9 n^5-661 n^4+7584 n^3+125232 n^2+465592 n+554064\right) \zeta_4}{4 (n+8)^8}
\w
+\frac{n \left(2 n^8-19 n^7+689 n^6+168914 n^5-416016 n^4-21086984 n^3-121746544 n^2-283766528 n-249483264\right) \zeta_3}{8 (n+8)^{10}}
\w
+ \Big(4 n^{10}+480 n^9+27419 n^8+921208 n^7+18509364 n^6+215607792 n^5+1332297632 n^4+4570604800 n^3
\w+8857566208 n^2+9208365056 n+4150108160\Big)\frac{n}{256 (n+8)^{11}}
\bigg]+\ca O (\epsilon^7)\ .
\label{phic}
\eea

\begin{figure*}[h]
\vspace*{-5mm}
\fboxsep0mm
\subfloat[]{\fig{.5}{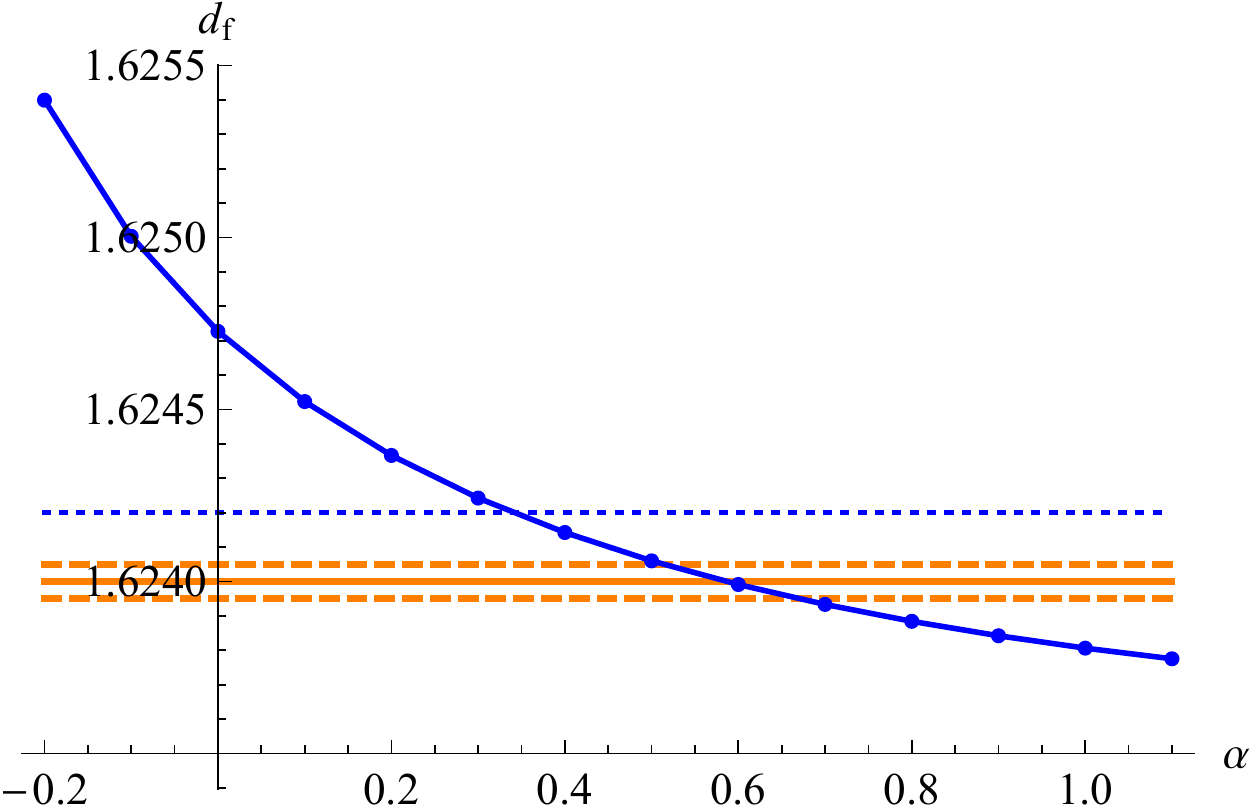}}~~
\subfloat[]{\fig{.5}{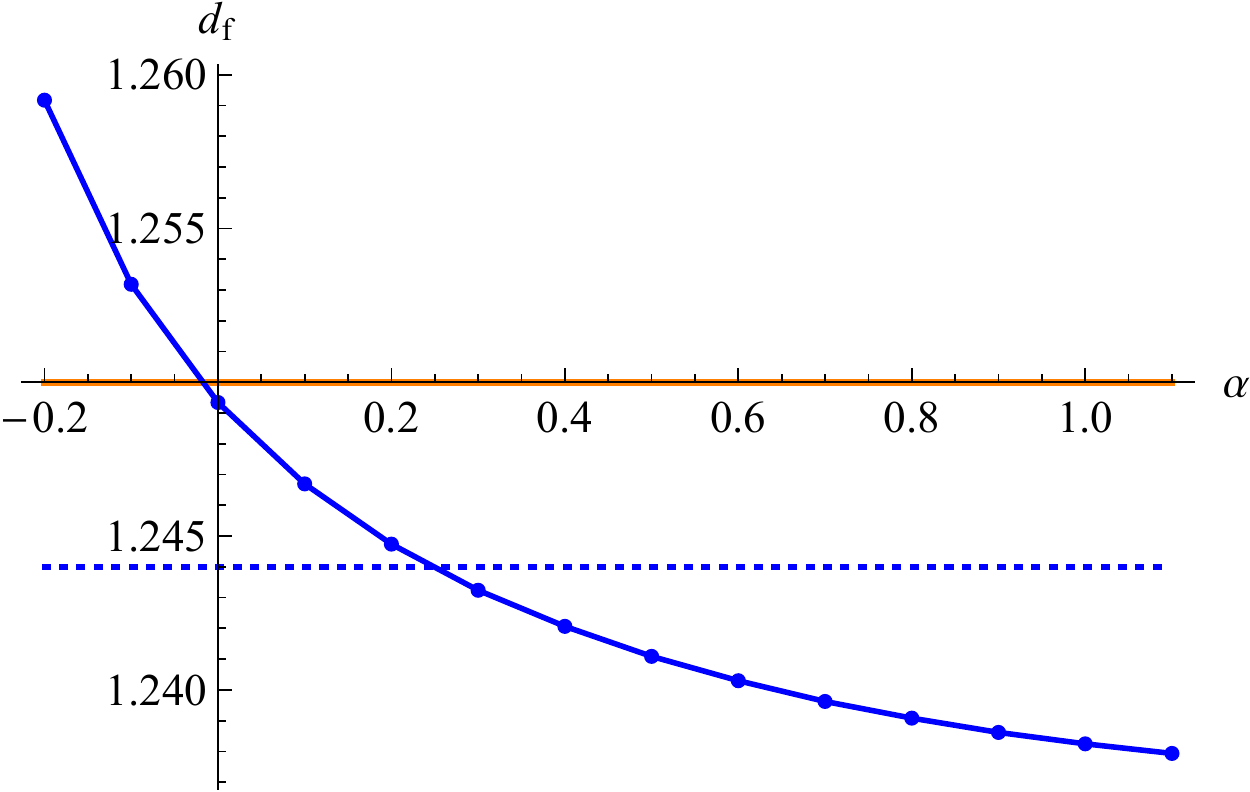}}
\caption{(a): In blue the fractal dimension  $d_{\rm f}$ of LERWs as a function of $\alpha$. The latter yields bounds for $d_{\rm f}$, i.e.\ $d_{\rm f}\in  [1.62378,1.6254]$, and as a best estimate the mean of the obtained values, $d_{\rm f}\approx 1.62426$ (blue dashed line). The numerical result is $d_{\rm f} = 1.624 00 \pm 0.00005$ (orange with error bars in dashes). \cite{Wilson2010}. (b): same for $d=2$. We find $d_{\rm f} \in [1.238,1.259]$, with a  mean estimate $d_{\rm f} = 1.244$, to be compared to the exact result $d_{\rm f}=5/4$.
Using only the 5-loop series gives $d_{\rm f}(d=3)\approx 1.621$, and $d_{\rm f}(d=2)=1.11.$
}
\label{f:z-alpha}
\end{figure*}

\end{widetext}

\begin{figure*}
\mbox{\begin{minipage}{8.6cm}\fig{1}{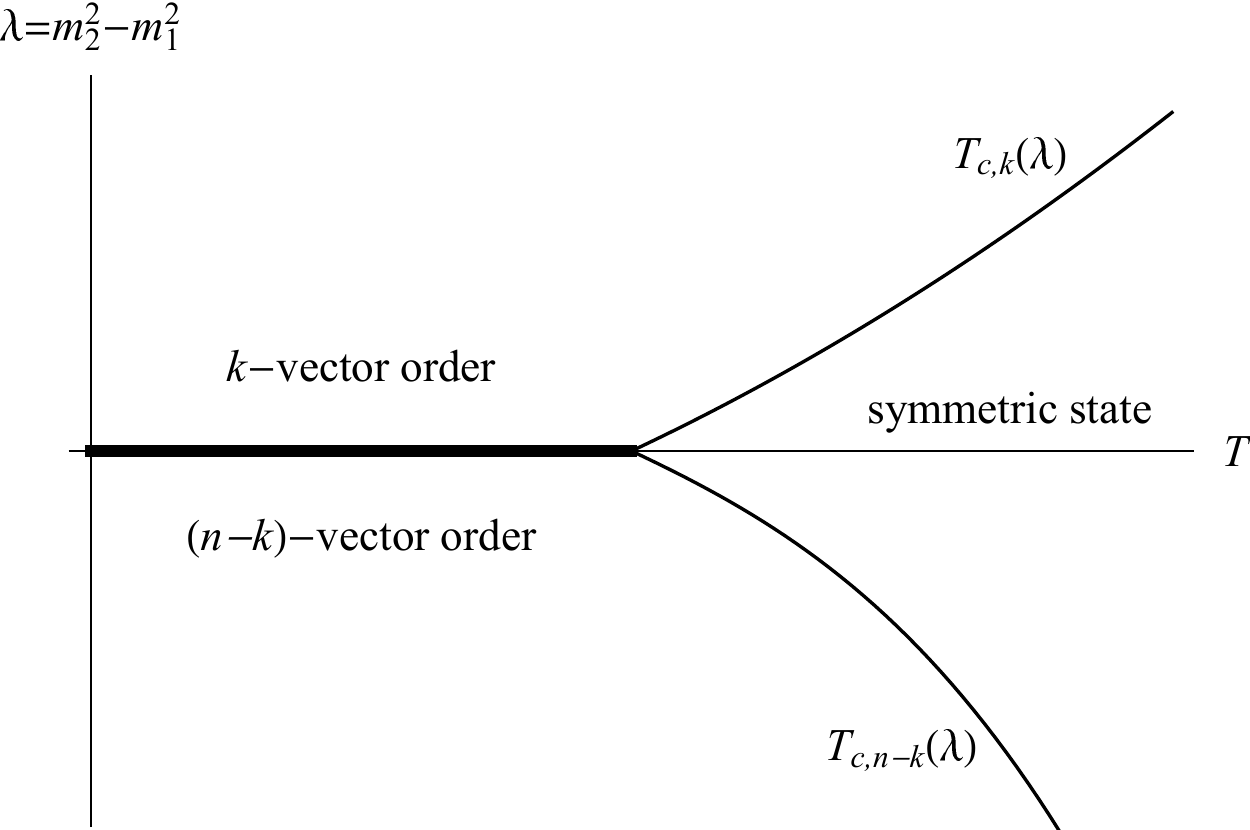}
\caption{The crossover phase  diagram  as given by  \cite{Amit}, with $\lambda = m_{2}^{2} -m_{1}^{2}$. The thick black line is a line of first-order phase transitions.}
\label{f:crossover-phase-diagram}
\end{minipage}}\hfill
\mbox{\begin{minipage}{8.6cm}\fig{1}{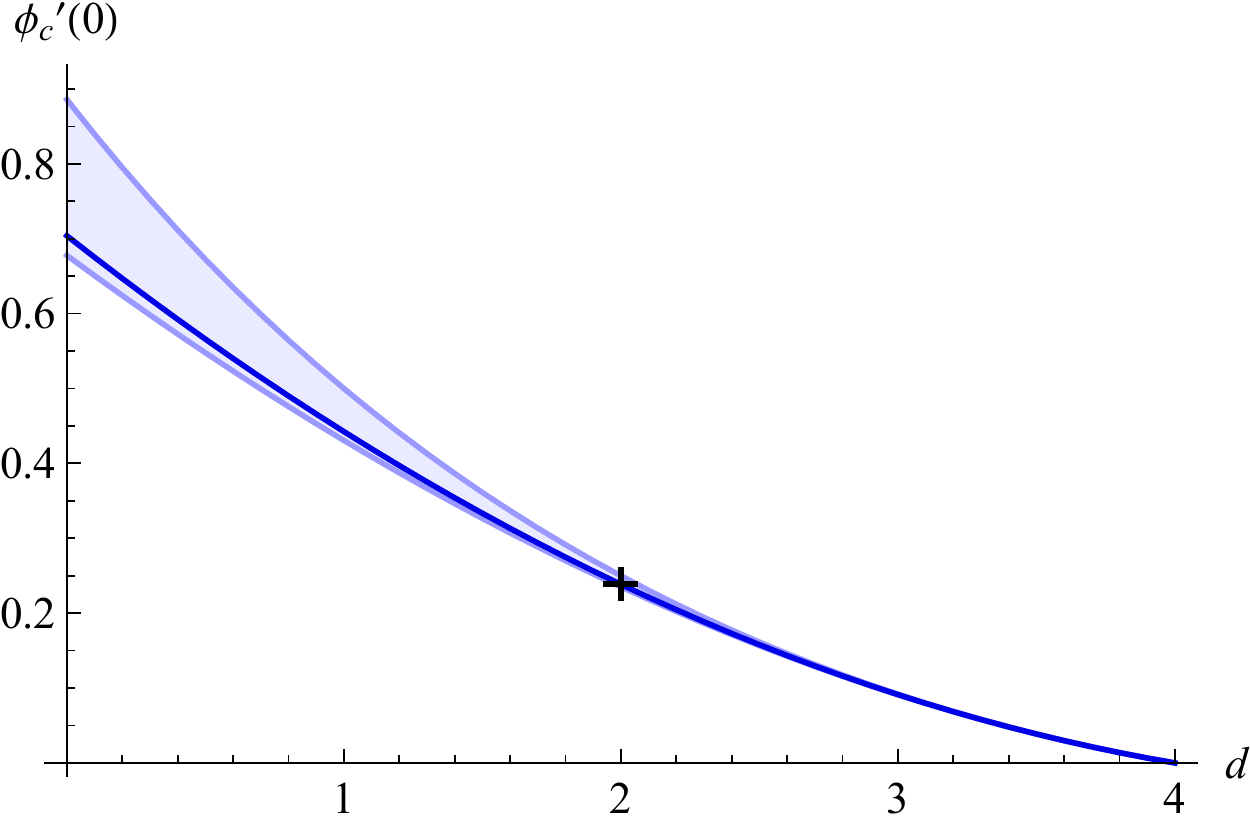}
\caption{Slope of the crossover exponent at $n=0$ for dimensions $0\le d\le 4$. The black cross is the analytic result from \Eq{phic-CFT} in $d=2$.}
\label{Fig9}
\end{minipage}}
\end{figure*}

This agrees with \cite{Kirkham1981} Eq.~(14) for $\phi_{\rm c}$ (noted $\phi$ there), except for a  misprint for the order $\epsilon^3$ term: the coefficient $682$ in the second line of Eq.~(14) of \cite{Kirkham1981} should read $628$.

The curve $\phi_{\rm c}(n)$, at least in higher dimensions is rather straight, thus the most important quantity to give is 
\bea
\phi_c'(0)|_{d=0} &=& 0.70(18) \\
\phi_c'(0)|_{d=1} &=& 0.44(6) \\
\phi_c'(0)|_{d=2} &=& 0.239(10) \\
\phi_c'(0)|_{d=3} &=& 0.0912(7) 
\ .
\eea
We have in all dimensions $d$
\be
\phi_c'(0) = \nu \left[ \gamma_{\tilde {\cal E}}'(0) -  \gamma_{ 1}'(0)\right] 
\ .
\ee
Estimates for $\phi_c'(0)$ obtained by self-consistent resummation (SC) and the procedure suggested in \cite{KompanietsPanzer2017} (KP17) are presented in table \ref{tab:dphic-numeric} and Fig.~\ref{Fig9}. Integrals of the inverse Borel transform do not converge well for $d=0$ in the KP17 resummation scheme, which prevents us to obtain an estimate there.

Explicit values for  the crossover exponent in $d=3$ to be compared with experiments, high-temperature series expansion and numerics are
\bea
\phi_{\rm c}^{\rm SC}(d=3,n=1)&=&1.089 (1) \ \\
\phi_{\rm c}^{\rm SC}(d=3,n=2)&=&1.180(4) \ \\
\phi_{\rm c}^{\rm SC}(d=3,n=3)&=&1.265(5) \  \\
\phi_{\rm c}^{\rm SC}(d=3,n=4)&=& 1.329(8)\  \\
\phi_{\rm c}^{\rm SC}(d=3,n=5)&=& 1.391(2)  \ .
\eea
There are   experiments for $n=2$ and $n=3$. For $n=2$:
\bea
\phi_{\rm c}^{\rm exp}(d=3,n=2)&=&1.17(2) ~~~\mbox{ \cite{RohrerGerber1977}}\\
\phi_{\rm c}^{\rm exp}(d=3,n=2)&=&1.18(5) ~~~\mbox{ \cite{Domann1979}} \\
\phi_{\rm c}^{\rm exp}(d=3,n=2)&=&1.23(4) ~~~\mbox{ \cite{MajkrzakAxeBruce1980}} \\
\phi_{\rm c}^{\rm exp}(d=3,n=2)&=&1.19(3) ~~~\mbox{ \cite{WalischPerez-MatoPetersson1989}} \\
\phi_{\rm c}^{\rm exp}(d=3,n=2)&=&1.17(10) ~\mbox{ \cite{WuYoungShaoGarlandBirgeneauHeppke1994}} \ .
\eea
The 
first paper \cite{RohrerGerber1977} examines the bicritical point in GdAl$\mathrm{O}_3$, and the second one \cite{Domann1979} the bicritical point in TbP$\mathrm{O}_4$. In the third~\cite{MajkrzakAxeBruce1980} the structural phase transition in K${}_2$SeO${}_4$ is investigated\footnote{This is the only experiment where the value of the crossover exponent is significantly higher than our (and other) estimates, but  its lower bound is  close to the theoretical values. The notation used in the experiments  is $\phi_{\rm c}=2-\alpha-\bar\beta$.}.
The fourth one \cite{WalischPerez-MatoPetersson1989} is related to a continuous phase transition in Rb${}_2$ZnCl${}_{4}$.
The last one is for the  nematic-smectic-A${}_{2}$ transition~\cite{WuYoungShaoGarlandBirgeneauHeppke1994}.

Let us proceed to $n=3$:
\bea
\phi_{\rm c}^{\rm exp}(d=3,n=3)&=&1.278(26) ~\mbox{ \cite{ShapiraOliveira1978}}\\
\phi_{\rm c}^{\rm exp}(d=3,n=3)&=&1.274(45)  ~\mbox{ \cite{ShapiraOliveira1978}}\\
\phi_{\rm c}^{\rm exp}(d=3,n=3)&=&1.279(31) ~\mbox{ \cite{KingRohrer1979}} \ .
\eea
The first two figures are for two different samples of the very nearly isotropic antiferromagnet RbMnF${}_{3}$~\cite{ShapiraOliveira1978}, the last one~\cite{KingRohrer1979} is for the bicritical point in MnF${}_{2}$.

In Ref.~\cite{Zhang1997} a theory based on ${\rm SO}(5)$, i.e.\ $n=5$, has been proposed to explain  superconductivity and antiferromagnetism
in a unified    model. While MC simulations support this scenario \cite{Hu2001,Hu2002}, it has been argued in  Ref.~\cite{Aharony2002} that the isotropic fixed point is unstable and breaks down into ${\rm SO}(2) \times {\rm SO}(3)$.

Recent Monte Carlo simulations~\cite{HasenbuschVicari2011} provide very precise estimates for the crossover exponent for $n=2,3,4$ (in terms of~\cite{HasenbuschVicari2011} $\phi_c=Y_2 \nu$):
\bea
\phi_{\rm c}^{\rm MC}(d=3,n=2)&=&1.1848(8) \\
\phi_{\rm c}^{\rm MC}(d=3,n=3)&=&1.2735(9)  \\
\phi_{\rm c}^{\rm MC}(d=3,n=4)&=&1.3567(15)
\eea  
The 
high-temperature series expansion of \cite{PfeutyJasnowFisher1974}
yields
\bea
\phi_{\rm c}^{\rm HT}(d=3,n=2)&=&1.175(15) \ ,\\
\phi_{\rm c}^{\rm HT}(d=3,n=3)&=&1.250(15) \ .
\eea
An alternative  to the $\epsilon$-expansion is to work directly in dimension $d=3$, (renormalization group in fixed space dimension $d=3$, denoted RG3), as was done in Ref.~\cite{CalabresePelissettoVicari2002}:
\bea
\phi_{c}^{\rm RG3}(n=2) &= &
1.184 (12) \\
\phi_{\rm c}^{\rm RG3}(n=3) &=&  1.271 (21) \\
 \phi_{\rm c}^{\rm RG3}(n=4) &=& 1.35 (4)   \\
 \phi_{\rm c}^{\rm RG3}(n=5) &=& 1.40 (4) \\
  \phi_{\rm c}^{\rm RG3}(n=8) &=&   1.55 (4) \\
   \phi_{\rm c}^{\rm RG3}(n=16) &=&    1.75 (6) \ .
\eea
Another approach is the non-perturbative  renormalization group (NPRG).  With this method the following estimates were obtained~\cite{EichhornMesterhazyScherer2013} (in terms of~\cite{EichhornMesterhazyScherer2013} $\phi_{\rm c} = \theta_1/\theta_2=y_{2,2}\nu$):
\bea
\phi_{\rm c}^{\rm NPRG}(n=2) &=&  1.209 \\
 \phi_{\rm c}^{\rm NPRG}(n=3) &=& 1.314    \\
 \phi_{\rm c}^{\rm NPRG}(n=4) &=& 1.407 \\
  \phi_{\rm c}^{\rm NPRG}(n=5) &=&   1.485  \\
   \phi_{\rm c}^{\rm NPRG}(n=10) &=&    1.710 \ .
\eea
Values provided by NPRG are systematically higher than those provided by other methods, but it is not clear how precise these values are. Their deviation from all other values is on the level of several percent, and we believe this to be an appropriate error estimate.

\begin{table*}
	\centering\caption{Numerical values for   $\phi_c(n)$ in $d=3$.}
\label{tab:phicd3-numeric}	\begin{tabular}{cccccccccc}
	\toprule
		$\quad n \quad$ & SC & SC from $\phi_{\rm c}^{-13/4}$ & KP17 & KP17  $\phi_{\rm c}^{-13/4}$& RG3~\cite{CalabresePelissettoVicari2002} & NPRG~\cite{EichhornMesterhazyScherer2013}& ~~~HT~\cite{PfeutyJasnowFisher1974}~~~ &~~~~ MC~\cite{HasenbuschVicari2011}~~~~ & experiment   \\
	\midrule
$2$ & $1.180(4)$ &1.183(1) & $1.183(3)$ & 1.1843(6) & $1.184(12)$& 1.209 & $1.175(15)$& $1.1848(8)$ &$1.17(2)$~~\cite{RohrerGerber1977}\\
 &  &  & & & &  & & &$1.18(5)$~~\cite{Domann1979}\\
 &  &  & & & & & & &$1.19(3)$~~{\cite{WalischPerez-MatoPetersson1989}}\\
 &  &  & & & & & & &$1.23(4)$~~{\cite{MajkrzakAxeBruce1980}}\\ 
 &  &  & & & & & & &$1.17(10)$~\cite{WuYoungShaoGarlandBirgeneauHeppke1994}\\
$3$ & $1.265(5)$ & 1.273(1) & $ 1.263(13)$ & 1.2742(10) & 1.271(21) & 1.314 & $1.250(15)$ &$1.2735(9)$ & $1.278(26)$~\cite{ShapiraOliveira1978}\\
  &  &  &  & & & &  &  &$1.274(45)$~\cite{ShapiraOliveira1978} \\
  &  &  &  & & & &  &  &$1.279(31)$~\cite{KingRohrer1979} \\
$4$ & 1.329(5)& 1.361(1)& $1.33(3)$& 1.3610(7) & $1.35(4)$ & 1.407 &  & {$1.3567(15)$} & \\
$5$ &1.391(2) & 1.442(2)& $1.42(4)$& 1.444(5) &$1.40(4)$ & 1.485 &  &  & \\
$8$ &1.534(2) & 1.64(1) & $1.59(7)$& 1.625(17) &$1.55(4)$ &  &  &  & \\
 	\bottomrule
	\end{tabular}\end{table*}
The most precise 6-loop estimates are obtained by a resummation of the $\phi_{\rm c}^{-13/4}$ expansion: they have lower error estimates (in both the S.C. and KP17 method) and   better agree  with the most precise values from Monte Carlo simulations. See also discussion in Sec. \ref{s:2d-resummation}.

A summary is provided in table \ref{tab:phicd3-numeric}.

\section{Loop-erased random walks}
\label{s:Loop-erased random walks}
\begin{figure}[b]
\fig{1}{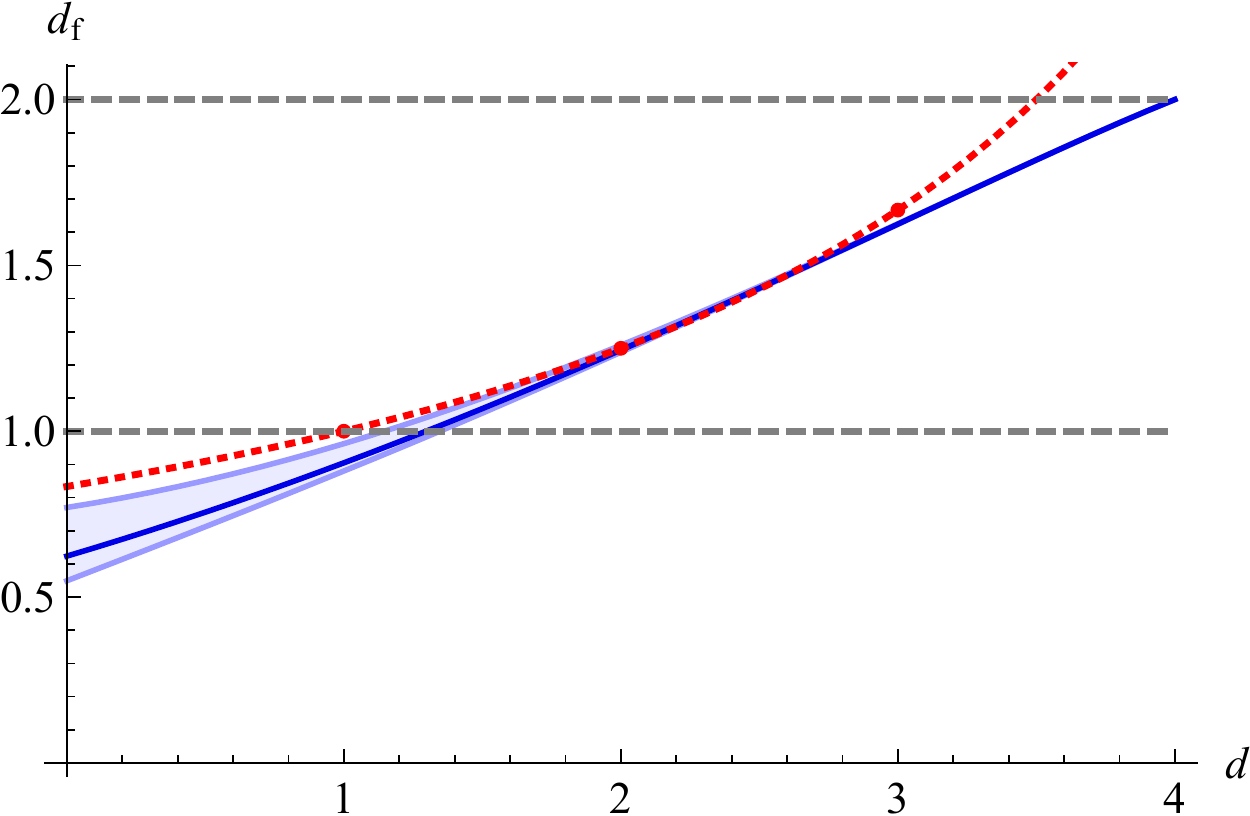}
\caption{$d_{\rm f}$ as a function of $d$ for LERW ($n=-2$). The red dashed line is the bound $d_{\rm f}\le \frac{5}{6-d}$ \cite{SapozhnikovShiraishi2018} (bound continuation to all dimensions guessed). The gray dashed lines are the bounds $1\le d_{\rm f}\le d_{{\rm RW}}=2$.}
\label{Fig10}
\end{figure}

The connection between the $O(n)$-symmetric $\phi^{4}$-theory at $n=-2$ and loop-erased random walks has only recently been established for all dimensions $d$ \cite{WieseFedorenko2018}, even though in $d=2$ this was known from integrability
\cite{Nienhuis1982,Duplantier1992}.  As we discussed above (see after Eq.~(\ref{df-SAW})), this is a random walk where loops are erased as soon as they are formed. As such it is a non-Markovian process.  On the other hand, its trace is equivalent to that of the {\em Laplacian Random Walk} \cite{LyklemaEvertszPietronero1986,Lawler2006}, which is Markovian, if one considers the whole trace as state variable. It is constructed on the lattice by solving the Laplace equation $\nabla^2 \Phi(x)=0$ with boundary conditions $\Phi(x)=0$ on the already constructed curve, while $\Phi(x)=1$ at the destination of the walk, either a chosen point, or infinity. The walk then advances from its tip $x$ to a neighboring point $y$, with probability proportional to $\Phi(y)$.
In dimension $d=2$, it is known via the relation to stochastic L\"owner evolution (SLE) \cite{LawlerSchrammWerner2004,Cardy2005} that the fractal dimension of LERWs is 
\be
d^{\rm LERW}_{\rm f}(d=2)=\frac 5 4\ .
\ee\begin{figure*}[t]
\subfloat[]{\fig{1}{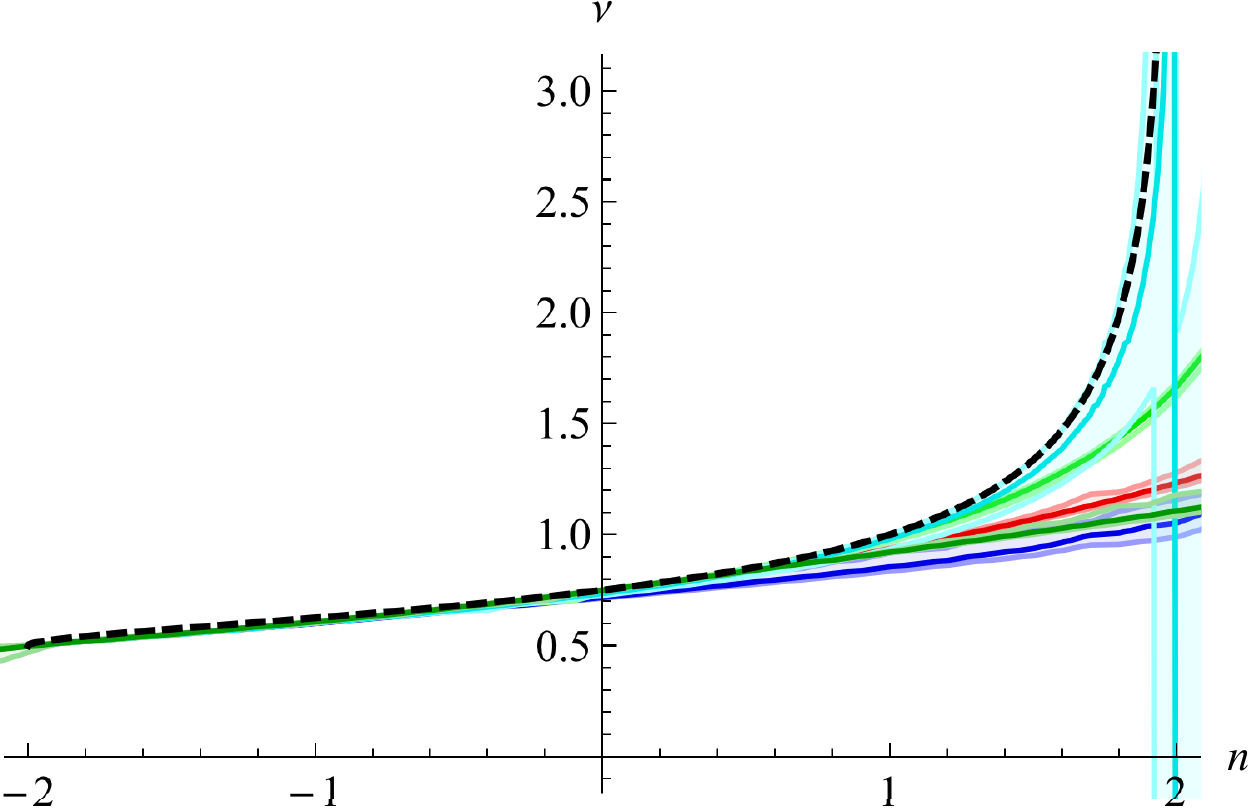}}\hfill
\subfloat[]{\fig{1}{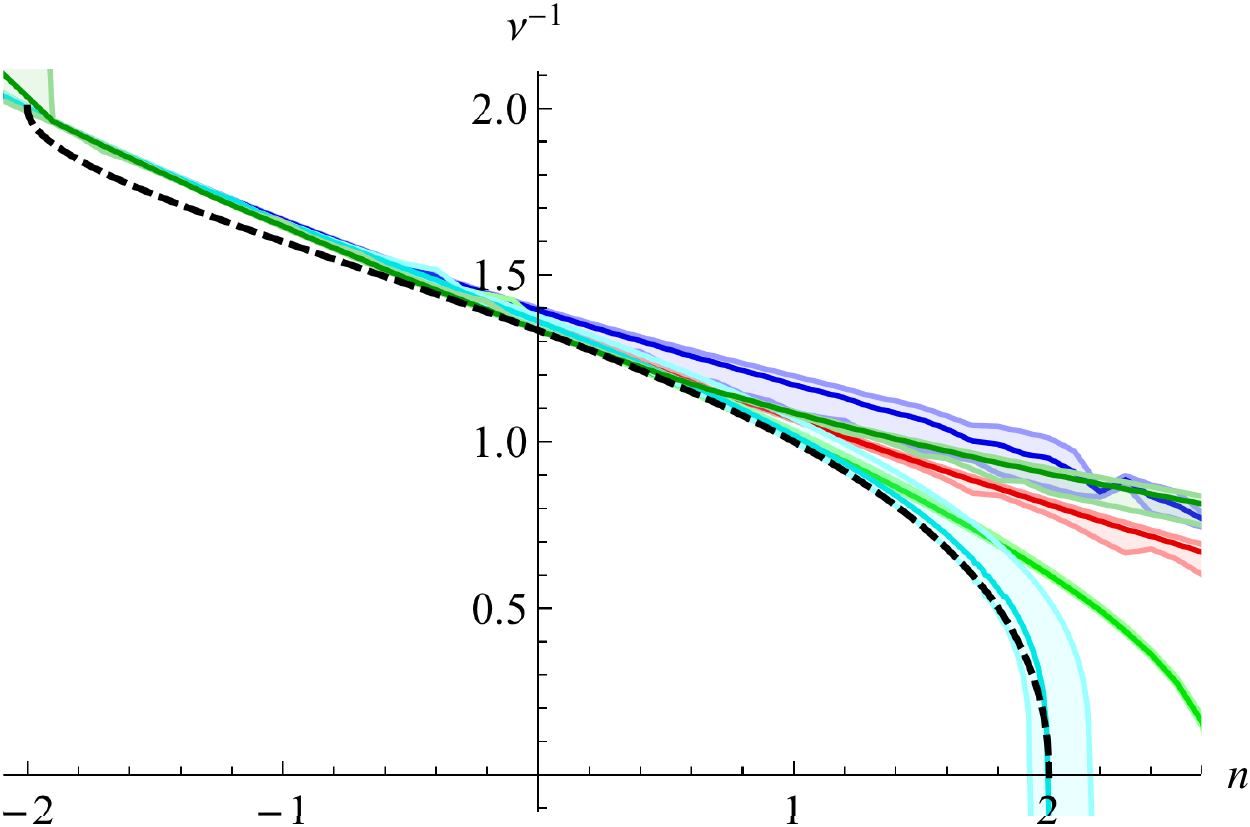}}
\caption{The exponent $\nu$ for $d=2$ (a) and its inverse (b). The different colors come from resummations of $\nu$ (blue), $1/\nu$ (red), $1/\nu^{2}$ (green), $1/\nu^{3}$ (cyan), and $\alpha=2-\nu d$ (dark green).  The dashed black line is  from CFT as given by Eq.~\eq{nu-CFT}. The shaded errors are (minimal) errors estimated from the uncertainty in the extrapolation, see section \ref{s:A self-consistent resummation procedure}.}
\label{f:nu:d=2}
\end{figure*}In three dimensions, there is no analytic prediction for the fractal dimension of LERWs, only the bound \cite{SapozhnikovShiraishi2018}
\be
1\le d_{\rm f}^{{\rm LERW}}\le \frac 53
\ .
\ee
We conjecture that it can be generalized to arbitrary dimension $d$ as 
\be
1\le d_{\rm f}^{{\rm LERW}}\le \frac 5{6-d}\ .
\ee
Note that this conjecture becomes exact in dimensions $d=1$ and $d=2$.
The best numerical estimation in $d=3$ is due to D.~Wilson \cite{Wilson2010}
\be
{d}_{\rm f, num}^{\rm LERW} (d=3) = 1.624 00 \pm 0.00005 = 1.62400(5)\ . 
\ee 
Our resummations   from the field theory  are (see Fig.~\ref{Fig10})
\bea
d_{\rm f, SC}^{\rm LERW}(d=3)&=& 1.6243 (10)\ .\nn\\
d_{\rm f, KP17}^{\rm LERW}(d=3)&=&1.623(6)\ .
\eea

\section{The limit of $d=2$ checked against conformal field theory}
\label{s:The limit of d=2 checked against conformal field theory}
\subsection{Relations from CFT}

In $d=2$, all critical exponents should be accessible via conformal field theory (CFT). The latter  is 
based on ideas proposed in the 80s by Belavin, Polyakov and Zamolodchikov \cite{BelavinPolyakovZamolodchikov1984}. They constructed a series of minimal models, indexed by an integer $m\ge 3$, starting with the Ising model at $m=3$. These models are conformally invariant and  unitary, equivalent to reflection positive in Euclidean theories. For details, see one of the many excellent textbooks on CFT \cite{DotsenkoCFT,DiFrancescoMathieuSenechal,ItzyksonDrouffe2,HenkelCFT}.
Their conformal charge is given by
\be
c=   1- \frac{6}{m(m+1)}\ .
\ee
The list of conformal dimensions allowed for a given $m$ is given by the Kac formula with integers $r,s$ (Eq.~(7.112) of \cite{DiFrancescoMathieuSenechal})
\be\label{36}
h_{r,s}= \frac{[r (m+1)-sm]^{2}-1}{4 m (m+1)} , \  1\le r < m, 1 \le s \le m. 
\ee
It was later realized that other values of $m$ also correspond to physical systems, in particular $m=1$ (loop-erased random walks), and $m=2$ (self-avoiding walks). These values  can further be extended to    the $\ca O(n)$-model with non-integer $n$ and $m$, using the identification
\be
n = 2 \cos\left(\frac \pi m\right)\ .
\ee
More strikingly, the table of dimensions allowed by \Eq{36} has to be extended to half-integer values, including $0$. 
It is instructive to read  \cite{JankeSchakel2010}, where all operators were identified.
This yields the fractal dimension of the propagator line \cite{RushkinBettelheimGruzbergWiegmann2007,BloeteKnopsNienhuis1992,JankeSchakel2010} \be\label{df-CFT}
d_{\rm f} = 2 -2 h_{1,0} = 1+ \frac{\pi }{2 \left(\arccos\left(\frac{n}{2}\right)+\pi
   \right)} \ .
\ee
This is compared to the $\epsilon $-expansion on Fig.~\ref{f:df:d=2}. 

For $\nu$, i.e.\ the inverse fractal dimension of all lines, be it propagator or loops, we get
\be\label{nu-CFT}
\nu =\frac1{2-2 h_{1,3}} = \frac14\left(1+\frac \pi{\arccos(\frac n2)}\right) \ .
\ee
This agrees with \cite{JankeSchakel2010}, inline after Eq.~(2). (Note that the choice $h_{2,1}$ coinciding with $h_{1,3}$  for Ising does nor work for general $n$.)  A comparison   to the $\epsilon $-expansion is given on Fig.~\ref{f:nu:d=2}.

For $\eta$, there are two suggestive candidates from the Ising model,  $\eta = 4 h_{1,2}= 4 h_{2,2}$. This does not work for other values of $n$.  
We propose in agreement with  \cite{RushkinBettelheimGruzbergWiegmann2007,BloeteKnopsNienhuis1992,JankeSchakel2010}
\be\label{eta-CFT}
\eta = 4 h_{\frac 1 2,0} = \frac{5}{4} -\frac{3 \arccos \left(\frac{n}{2}\right)}{4 \pi
   }-\frac{\pi }{\arccos  \left(\frac{n}{2}\right)+\pi
   }\ .
\ee
It has a square-root singularity both for $n=-2$ and $n=2$. A comparison to field theory is given on Fig.~\ref{f:eta:d=2}.

As we discuss in the next section, we have no clear candidate for the exponent $\omega$. {This is apparent on Fig.~\ref{f:omeag:d=2}, where our estimates from the resummation are confronted to some guesses from CFT. }

Finally the crossover exponent $\phi_{\rm c}$ defined in Eqs.~\eq{38} and \eq{54} becomes
\be\label{phic-CFT}
\phi_{\rm c} = \nu d_{\rm f} = \frac{1-h_{1,0}}{1-h_{1,3}} = \frac14 + \frac {3\pi}{8 \arccos (\frac n2)}\ .
\ee
This is compared to the $\epsilon $-expansion on Fig.~\ref{f:phic:d=2}.

\subsection{Resummation}
\label{s:2d-resummation}

\begin{figure}[t]

\fig{1}{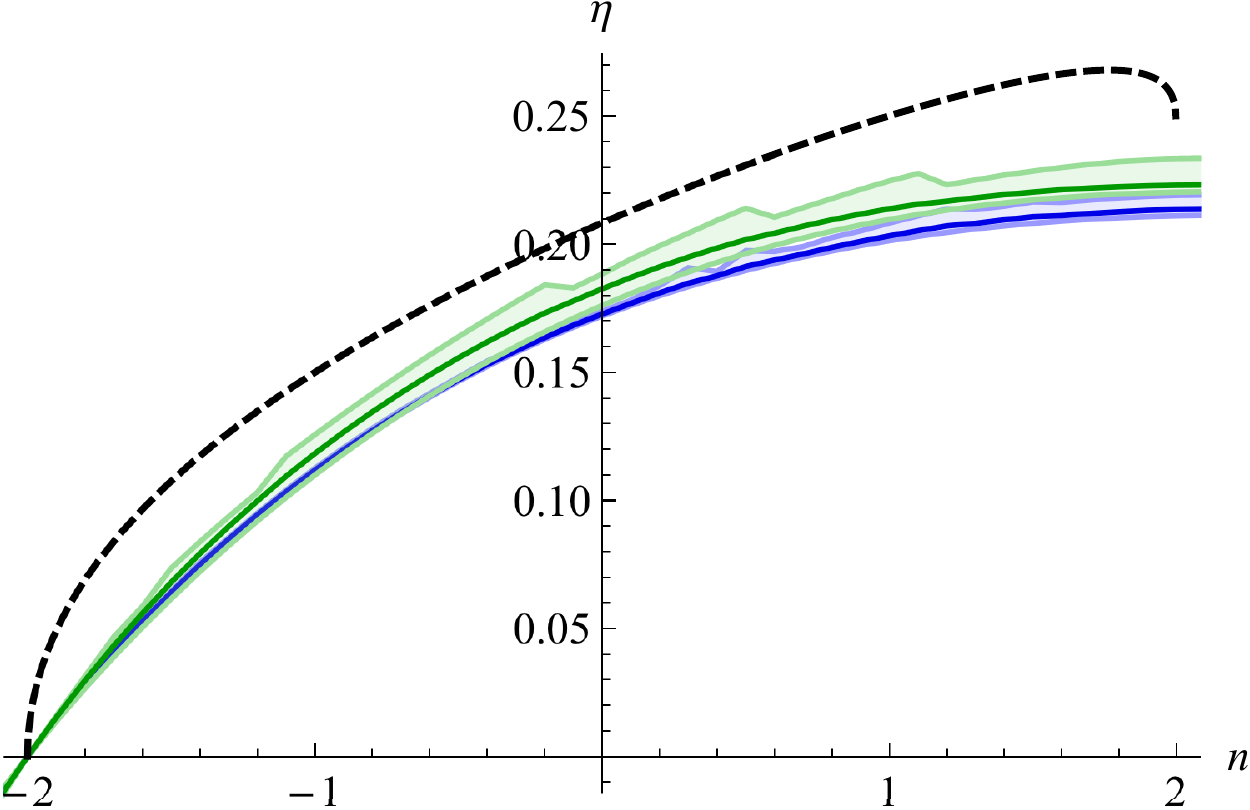}
\caption{The exponent $\eta$   in $d=2$. The blue curve is the direct expansion, the green one a resummation of $\sqrt{\eta}$, which, as $\eta$ starts at order $\epsilon^{2}$, has a regular series expansion in $\epsilon$.  
The black solid line  is $\eta=4 h_{1/2,0}$ as given by  \Eq{eta-CFT}.}
\label{f:eta:d=2}

\end{figure}

Note that there are singularities at $n=\pm 2$, the most severe one being the one at $n=2$ for the exponent $\nu$. For this reason, resummation is difficult for $n\approx 2$. We found that the singularity in $d=2$ is much better  reproduced when resumming $1/\nu^{3}$ instead of $\nu$, see Fig.~\ref{f:nu:d=2}. This expansion catches the divergence at $n=2$ in $d=2$, even though the singularity thus constructed is not proportional to $1/\sqrt{2-n}$, but proportional to $1/\sqrt[3]{2-n}$. As we will see below, reproducing this singularity at least approximately renders expansions also  more precise  in $d=3$, even for $n=0,1$.

The same situation appears for $\phi_{\rm c}$, where $1/\phi_{\rm c}^{13/4}$ provides the most precise fit of the $n=2$ singularity (see Fig.~\ref{f:phic:d=2}).   This leads to smaller error bars for both resummation methods (see Table ~\ref{tab:phicd3-numeric}) and supports our statement about the necessity of a proper choice of the object for resummation,  based on the knowledge of the $d=2$ singularities.

For $d_{\rm f}$ (Fig.~\ref{f:df:d=2}) and $\eta$ (Fig.~\ref{f:eta:d=2}),  the $\epsilon$-expansion is approximately correct. But there are square-root singularities when approaching $n=\pm2$ in $d=2$, which are not visible in the $\epsilon$-expansion. It is suggestive that these singularities   in $d=2$  influence the convergence in $d=3$. Building in these exact results in $d=2$, including the type of singularity in the $(d,n)$-plane  would increase significantly the precision in $d=3$. 

\begin{figure}[t]
\fig{1}{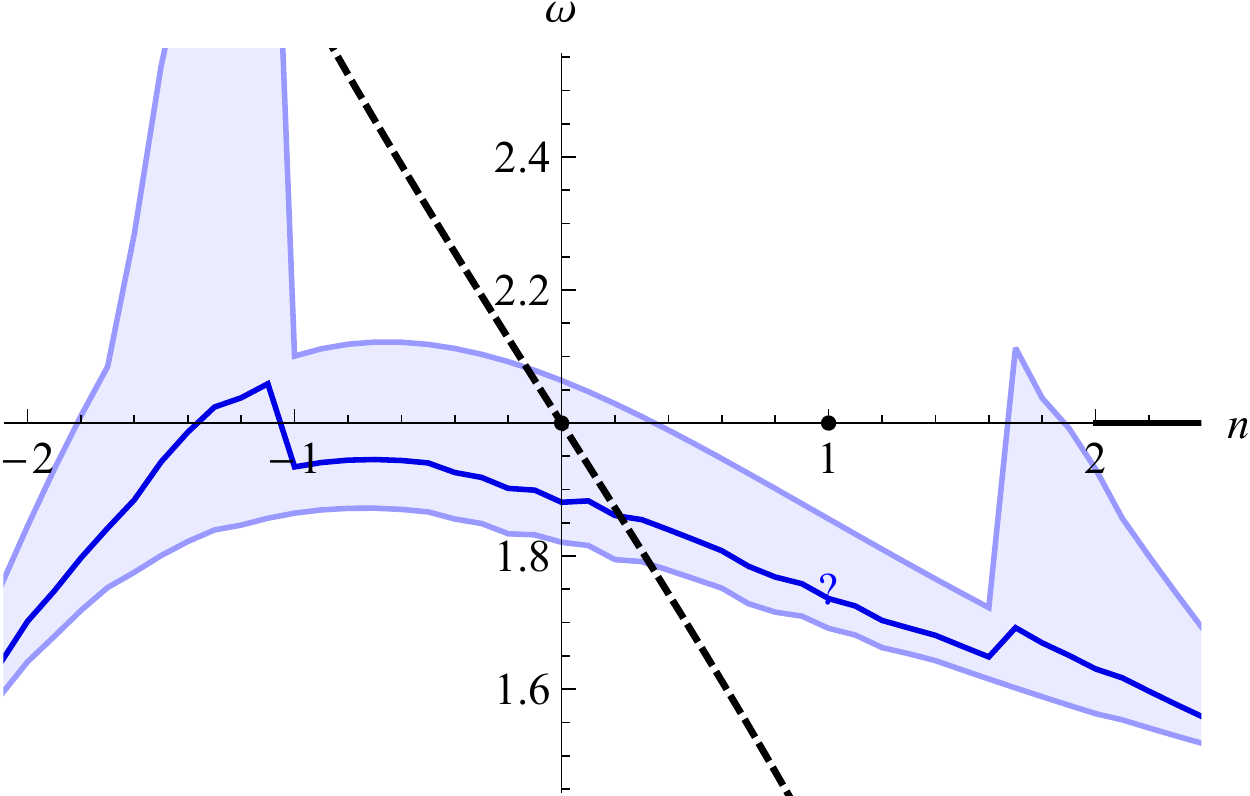}
\caption{The exponent $\omega$ in $d=2$.  Dots represent     values reported in the literature, mostly based on CFT. The value $\omega =7/4$ for $n=1$ is consistent with the $O(1)$-term in \cite{BarouchMcCoyWu1973}, while   the reanalysis of  \cite{CalabreseCaselleCeliPelissettoVicari2000}   concludes on $\omega=2$. \cite{CalabreseCaselleCeliPelissettoVicari2000} also argue that $\omega=2$ for $n>2$. The black dashed line is the guess \eq{omega:guess:d=2} resulting from the operator generating an intersection between two lines.}
\label{f:omeag:d=2}
\end{figure}

\begin{figure}[t]
\fig{1}{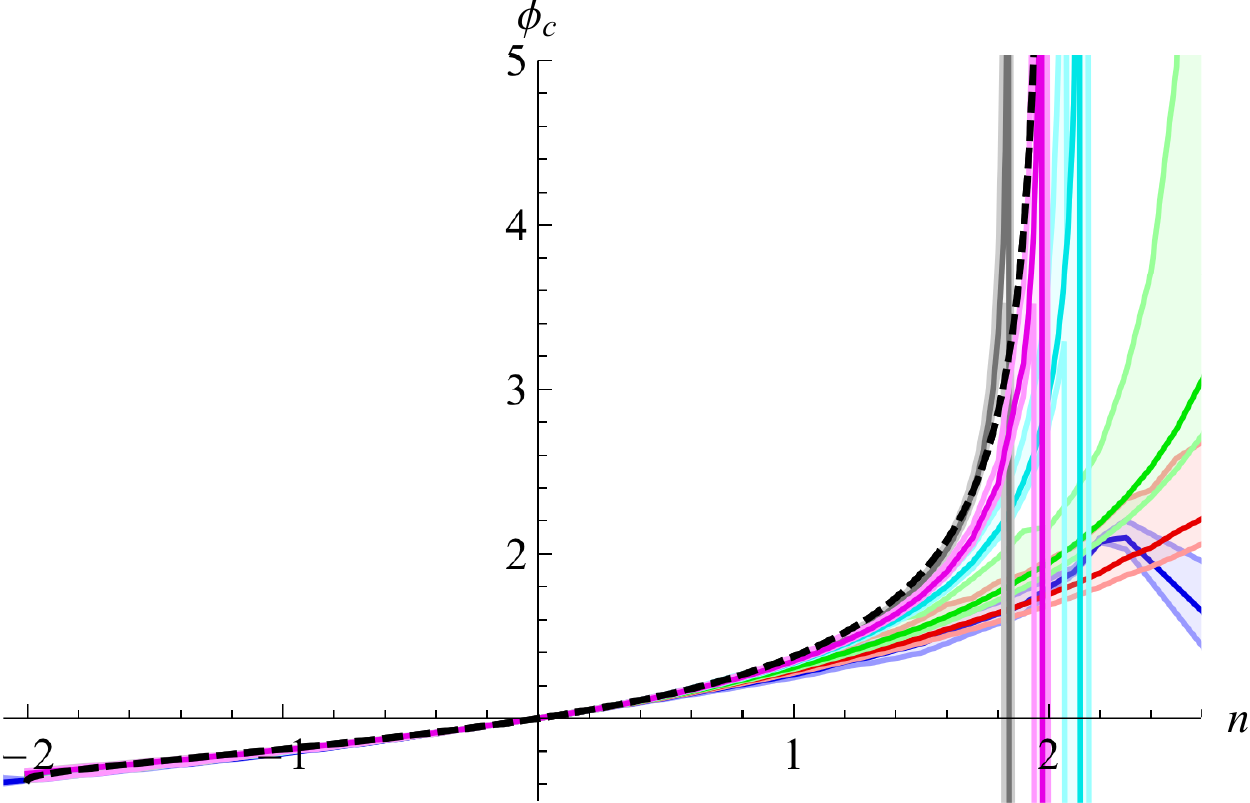}
\caption{The exponent $\phi_{\rm c}$  in $d=2$. The dashed black line is the analytic result from \Eq{phic-CFT}. The colored lines are resummations of $\phi_{\rm c}$ (blue), $1/\phi_{\rm c}$ (red), $1/\phi_{\rm c}^2$ (green), $1/\phi_{\rm c}^3$ (cyan), $1/\phi_{\rm c}^{13/4}$ (magenta), and $1/\phi_{\rm c}^{3}$ (gray). Resumming $1/\phi_{\rm c}^{13/4}$ considerably improves the precision.}
\label{f:phic:d=2}
\end{figure}

As for $\omega$ presented on Fig.~\ref{f:omeag:d=2}, the situation is rather unclear, as there is no choice of ${h_{r,s}}$ which is a good  candidate for all $n$ in the range of
$-2\le n \le 2$. 
Intersections in high-temperature graphs are given by $h_{2,0}$, and this operator is the closest  in spirit to the  $(\vec \phi^{2})^{2}$-interaction of our field theory, resulting into 
\be\label{omega:guess:d=2}
\omega_{\rm guess} = 2 h_{2,0}-2\ .
\ee  
This contradicts the results from the $\epsilon$-expansion presented on Fig.~\ref{f:omeag:d=2}.
It is not even clear whether this is a question which can be answered via CFT: As all observables depend on the coupling $g$, the exponent $\omega$ quantifies how far this coupling has   flown to the IR fixed point. On the other hand, in a CFT the ratio of size $L$ over lattice cutoff $a$ has gone to infinity, and the theory by construction is at $g=g_*$. Our results are consistent with  $\omega = 2$ for all $n$, in which case the associated operator might simply be the determinant of the stress-energy tensor, sometimes (abusively) referred to as   $T\bar T$, see e.g.\  \cite{Cardy2018}.

\begin{figure*}[t]
\fboxsep0mm\mbox{\begin{minipage}{8.6cm}\fig{1}{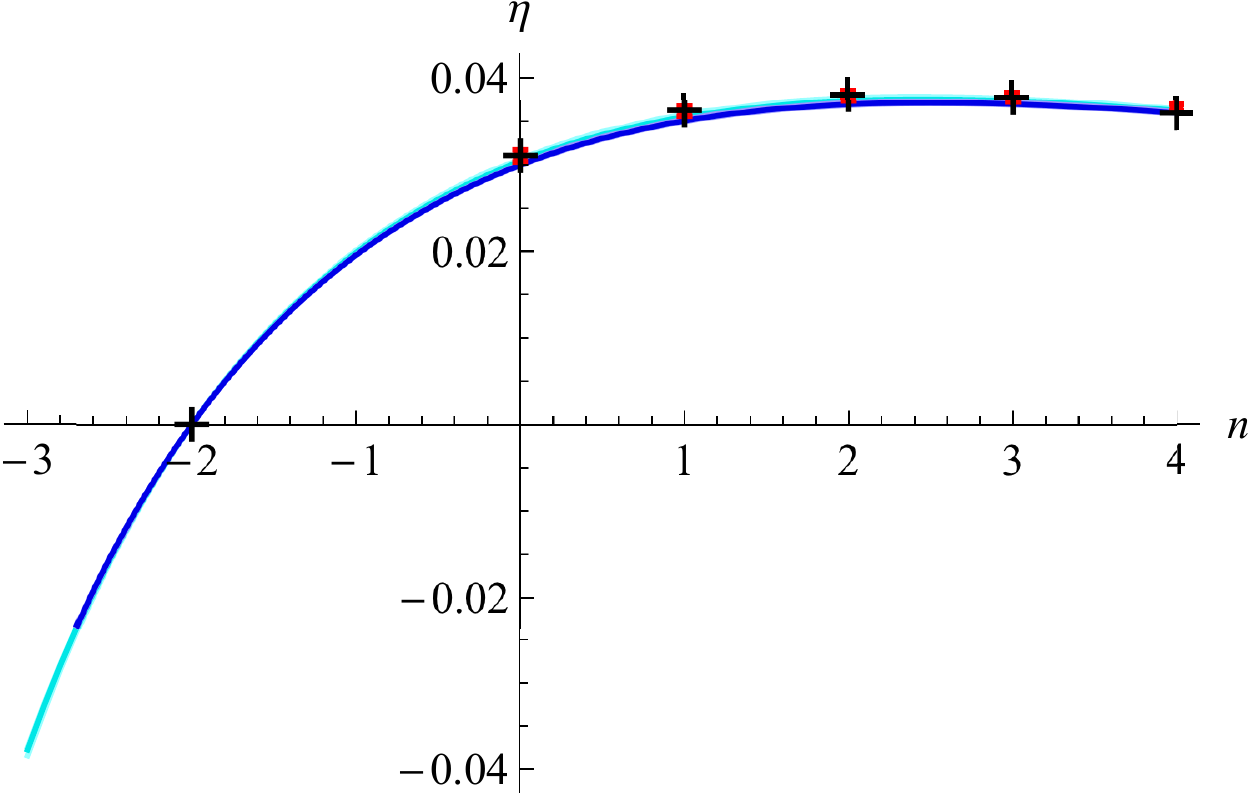}
\caption{The exponent $\eta$ in $d=3$.   The SC resummation scheme (in blue) seems to be systematically   smaller than the values of KP17 (in red). SC resummation of $\sqrt{\eta}$ (in cyan) works slightly better. Black crosses represent the best values from MC and conformal bootstrap, as  given in \cite{KompanietsPanzer2017}.}
\label{f:eta:d=3}
\fig{1}{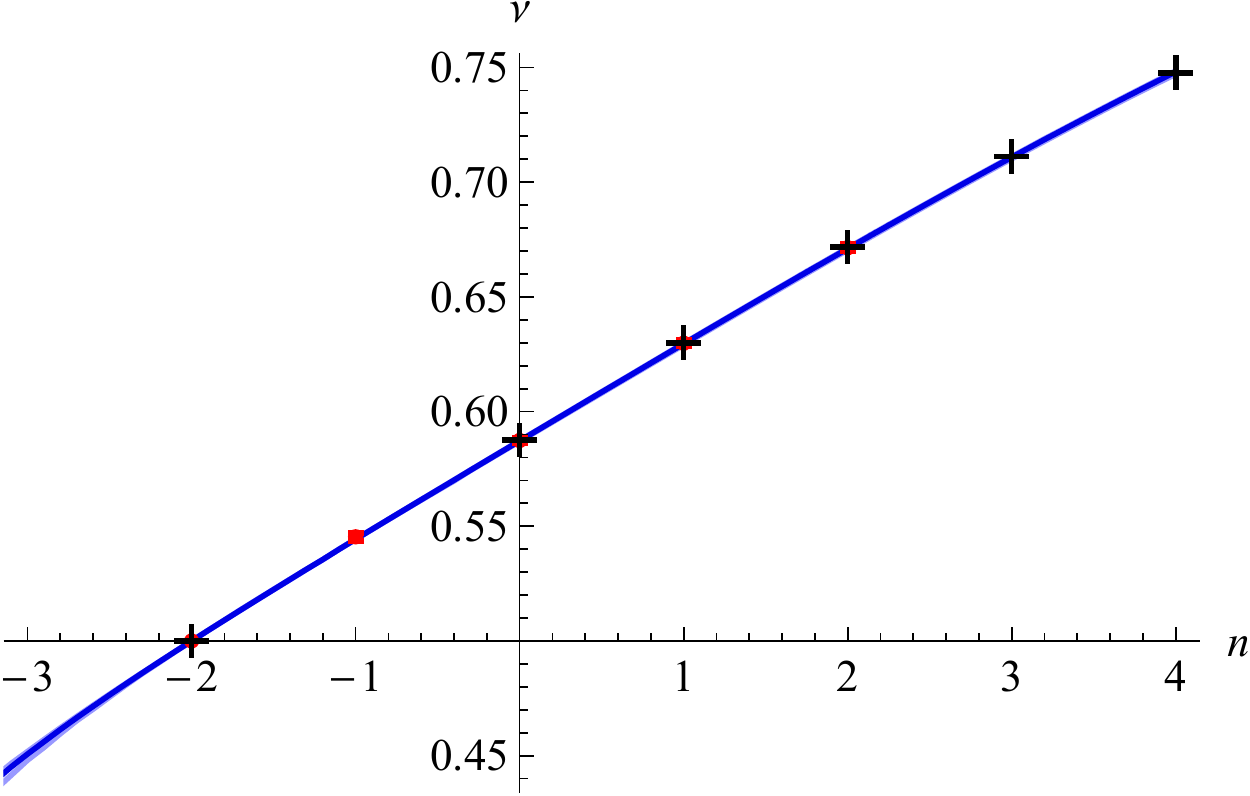}
\caption{The exponent $\nu$ in $d=3$, obtained from a resummation of $1/\nu^{3}$. In blue the results from   SC, in red using KP17. Black crosses  are from  MC and conformal bootstrap, as given in \cite{KompanietsPanzer2017}.}
\label{f:nu:d=3}
\end{minipage}\noindent~~~~~~~~
\begin{minipage}{8.6cm}
\fig{1}{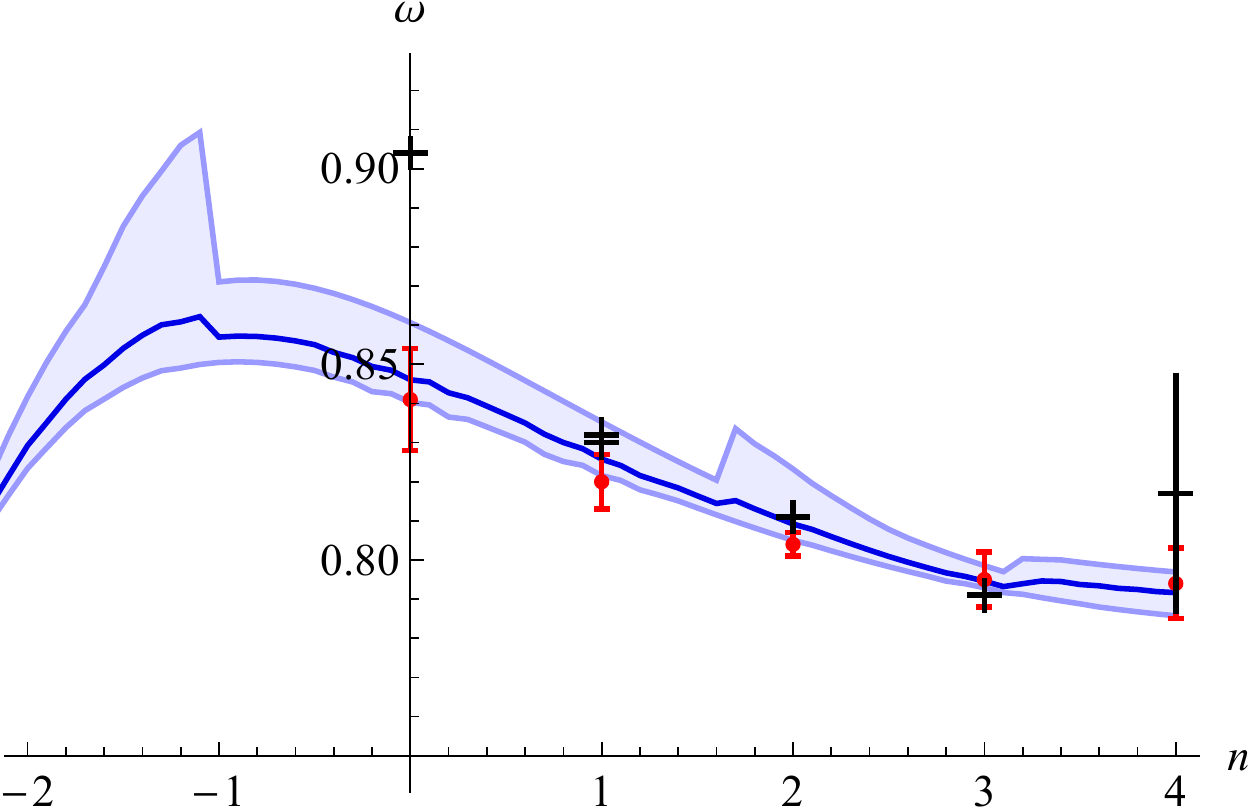}
\caption{The exponent $\omega$ in $d=3$  via SC (blue, with shaded error bars), and  KP17 (in red). Crosses represent the best values from MC, as given in Ref.~\cite{KompanietsPanzer2017}.}
\label{f:omega:d=3}
\fig{1}{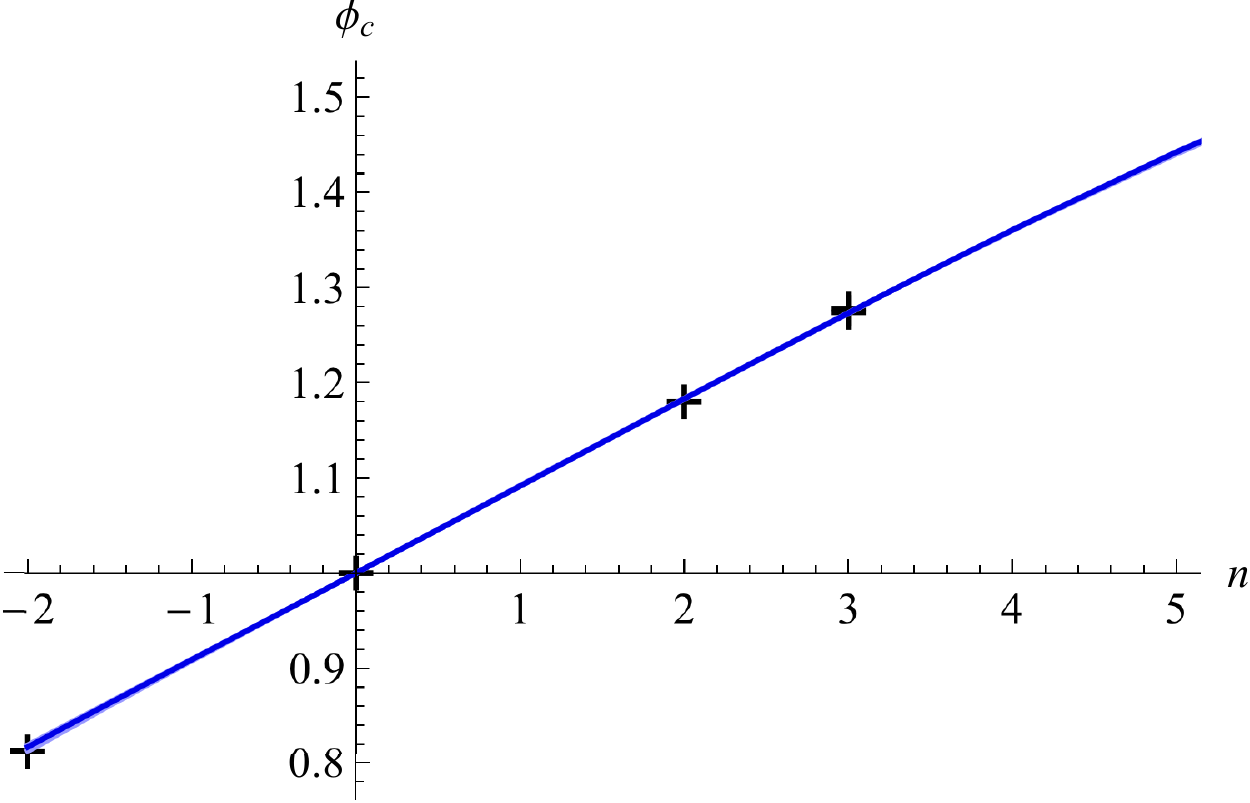}
\caption{The exponent $\phi_{\rm c}$  in $d=3$. Crosses are from MC and experiments \cite{Domann1979,ShapiraOliveira1978}. The value for $n=-2$ is taken as $d_{\rm f}/2$ with $d_{\rm f}$ the fractal dimension of LERWs \cite{Wilson2010}.}
\label{f:phic:d=3}
\end{minipage}}

\end{figure*}

\section{Improved estimates in $d=3$ for all exponents}
\label{s:estimates in d=3}
With the knowledge gained in $d=2$, we are now in a position to give our best estimates for all critical exponents. For the exponent $\nu$, we use the expansion of $1/\nu^3$, while for $\eta$  and $\omega$ we use the standard direct expansions. For $d_{\rm f}$ we both use the direct expansion, as the expansion of $1/d_{\rm f}$, to get an idea about the errors induced by changing the quantity to be extrapolated. 

Our findings are given on Tables \ref{tab:phicd3-numeric}  to \ref{tab:omega-numeric} as  well as  Figs.~\ref{tab:df-d=3} and  \ref{f:eta:d=3} to \ref{f:phic:d=3}.
Let us  summarize them:  

The exponent $\eta$ is shown on Table~\ref{tab:eta-numeric} and figure \ref{f:eta:d=3}. For SAWs, the agreement of KP17 with the Monte-Carlo results of 
\cite{Clisby2017,ClisbyDunweg2016} is  better than $10^{-3}$ (relative). For the Ising model ($n=1$),  the agreement  with the conformal bootstrap \cite{KosPolandSimmons-DuffinVichi2016} is of the same order.

Our predictions for $\nu$ are given on table \ref{tab:nu-numeric} and Fig.~\ref{f:nu:d=3}. Using the expansion of $1/\nu^3$, the relative deviation to the conformal bootstrap is about $3\times 10^{-4}$ instead of $10^{-3}$ for the direct expansion, validating both schemes. The same deviation of $3\times 10^{-4}$  appears in the comparison to Monte Carlo simulations of SAWs.

The exponent $\phi_{\rm c}$ has already been discussed in section \ref{s:crossover}. Table \ref{tab:phicd3-numeric} summarizes our findings. In general, there is a very good agreement between the diverse theoretical predictions and experiments. We find it quite amazing that experiments were able to measure this exponent with such   precision.  

Via the relation \eq{54}, which can be written as 
$\phi_{\rm c} =\nu d_{\rm f} $, the exponent $\phi_{\rm c}$ is intimately related to the fractal dimension $d_{\rm f}$ of curves discussed in the introduction, and summarized on Fig.~\ref{tab:df-d=3}. Again, in all cases the agreement is well within the small error bars.

The exponent $\omega$ is notoriously difficult to obtain, possibly due to a non-analyticity of the $\beta$-function at the fixed point $g_{*}$ \cite{CalabreseCaselleCeliPelissettoVicari2000}. We show our predictions on table \ref{tab:omega-numeric} and Fig.~\ref{f:omega:d=3}. The deviations from results obtained by other methods are much larger, but consistent with our error bars.   The only  value from simulations we have doubts about is $\omega$ for SAWs in $d=3$, which is an ``outsider'' on  Fig.~\ref{f:omega:d=3}. As reported by \cite{Belohorec1997,ClisbyDunweg2016},   
\bea
\omega= \Delta /\nu &=&0.899(14)\qquad \mbox{\cite{ClisbyDunweg2016}}\ ,\\
\omega=\Delta /\nu &=&0.904(6)\qquad~~ \mbox{\cite{Belohorec1997}} \ .
\label{105}
\eea
Ref.~\cite{ClisbyDunweg2016} provides the   most precise result for $\nu=0.58759700(40)$, while the value of $\Delta=\omega\nu=0.528(8)$ is less precise than that of Ref.~\cite{Belohorec1997}, namely $\Delta=0.5310(33)$.
The value    $\nu= 0.58756(5)$ of Ref.~\cite{Belohorec1997} is less precise than the one of Ref.~\cite{ClisbyDunweg2016}, but the error is negligible   compared to that of $\Delta$. 
Combining the most precise values gives an estimate $\omega=0.904(5)$ as in \Eq{105}, but with a slightly reduced error bar.

As already stated, 
proper choice of the object of resummation can significantly increase the convergence and yield estimates   closer to those of CFT in $d=2$, and conformal bootstrap in $d=3$. While for the exponent $\nu$ this choice is   obviously   $\nu^{-3}$, and for $\phi_{\rm c}$ it is $1/\phi_{\rm c}^{13/4}$,  since both catch   the singularity in  $d=2$ (see Figs.\ \ref{f:nu:d=2} and \ref{f:phic:d=2}), for the exponents $\eta$ and $\omega$ there is no evident choice.
A more detailed investigation of these ideas is beyond the scope of the present paper, and left for future research.

\begin{table}
\centering\caption{Numerical values for the exponent $\eta$ in $d=3$. SC combines expansion for $\eta$ and $\sqrt \eta$.}\label{tab:eta-numeric}	\begin{tabular}{cccc}
	\toprule
		$\quad n \quad$ & SC & KP17 & other  \\
	\midrule
		-2 & 0 & 0 & 0\\
		-1 & 0.0198(3) & 0.0203(5) &\\
		0 & 0.0304(2)& 0.0310(7) \cite{KompanietsPanzer2017}& 0.031043(3) \cite{Clisby2017,ClisbyDunweg2016}\\
		1 & 0.0355(3)&0.0362(6) \cite{KompanietsPanzer2017}& 0.036298(2) \cite{KosPolandSimmons-DuffinVichi2016}\\
		2 & 0.0374(3)&0.0380(6) \cite{KompanietsPanzer2017}&0.0381(2) \cite{CampostriniHasenbuschPelissettoVicari2006}\\
		3 & 0.0373(3)&0.0378(5) \cite{KompanietsPanzer2017}&0.0378(3) \cite{HasenbuschVicari2011}\\
		4 & 0.0363(2)  &0.0366(4) \cite{KompanietsPanzer2017}&0.0360(3) \cite{HasenbuschVicari2011}\\
 	\bottomrule
	\end{tabular}\end{table}
\begin{table}[t]
	\centering\caption{Numerical values for the exponent $\nu$  in $d=3$.}	\label{tab:nu-numeric}	\begin{tabular}{ccccc}
	\toprule
		$\quad n \quad$ & SC ($\nu^{-3}$) & KP17 ($\nu^{-3}$)& KP17 ($1/\nu$) & other  \\
\midrule
		-2 & 0.5 & 0.5&0.5&\\
		-1 & 0.54436(2) &0.545(2) & 0.5444(2)\\
		0 & 0.5874(2) & 0.5874(10) & 0.5874(3) \cite{KompanietsPanzer2017} &0.5875970(4) \cite{ClisbyDunweg2016}\\
		1 &0.6296(3) &0.6298(13) & 0.6292(5) \cite{KompanietsPanzer2017}& 0.629971(4) \cite{KosPolandSimmons-DuffinVichi2016}\\
		2 &0.6706(2) &0.6714(16)& 0.6690(10) \cite{KompanietsPanzer2017}&0.6717(1) \cite{CampostriniHasenbuschPelissettoVicari2006}\\
		3 &0.70944(2)& 0.711(2) &0.7059(20) \cite{KompanietsPanzer2017}& 0.7112(5) \cite{CampostriniHasenbuschPelissettoRossiVicari2002}\\
		4 & 0.7449(4) & 0.748(3)&0.7397(35) \cite{KompanietsPanzer2017}& 0.7477(8) \cite{Deng2006}\\
 	\bottomrule
	\end{tabular}	\vspace*{-2mm}
\end{table}

\section{Connection to the large-$n$ expansion}
\label{s:Connectiontolarge-Nexpansion}

One of the most effective checks of perturbative expansions is comparison of   different expansions of the same quantity. For the $O(n)$ model, the $\epsilon$-expansion provides a  series in $\epsilon$ which is an   exact function in $n$, while the large-$n$ expansion (or $1/n$-expansion) provides a series in $1/n$ with coefficients exact in $d$. Thus setting $d=4-\epsilon$ in the  $1/n$ expansion and expanding it in $\epsilon$, while  expanding the coefficients  of the $\epsilon$-expansion in $1/n$ for the same quantity  must yield identical 
series. As for  each expansion a  different method is  used,  this provides a very strong cross check for both expansions.

The large-$n$ expansion of the crossover exponent $\phi_c$ as given in Eqs.~\eqref{38} and \eqref{54} was calculated in  \cite{Gracey2002} up to  $1/n^2$.  Expanding it in $\epsilon$ we obtain a double $(\epsilon, 1/n)$-expansion for  $ \phi_c^{(\epsilon,n)}$,  
\bea
&&
\phi_c^{(\epsilon,n)}=\Big[1+\frac{\epsilon }{2}+\frac{\epsilon
   ^2}{4}+\frac{\epsilon
   ^3}{8}+\frac{\epsilon
   ^4}{16}+\frac{\epsilon
   ^5}{32}+\frac{\epsilon ^6}{64} + \ca O(\epsilon^7)\Big]\nn
\\
&&   +\frac{1}{n}\Big[
   -4 \epsilon +\epsilon ^3
   +\big(-\zeta_3+1\big)\epsilon^4
   +\frac{3 }{4}\big(- 
   \zeta_4+1\big)\epsilon
   ^5
\w
 \qquad ~ +\frac{1}{4}\big(-3\zeta_5+\zeta_3+2\big)\epsilon^6
   + \ca O(\epsilon^7)\Big]\nn
\\
&&   +\frac{1}{n^2}\Bigg[
   32 \epsilon
   -31 \epsilon ^2
   - \Big(30\zeta_3-\frac{43}{2}\Big)\epsilon ^3
 \w \qquad ~    +\Big(40 
   \zeta_5-\frac{45  \zeta_4}{2}+61
    \zeta_3-\frac{155}{16}\Big)\epsilon ^4
   \w
  \qquad ~   +\Big(50\zeta_6
    +8 \zeta_3^2
    -115 \zeta_5
	+\frac{183 \zeta_4}{4}
   -9 \zeta_3
    +\frac{61 }{16}\Big)\epsilon^5
    \w
   \qquad ~  +\Big(
   71 \zeta_7
   +12 \zeta_3 \zeta_4
   -\frac{1075 \zeta_6}{8}
   -35 \zeta_3^2
   +\frac{195 \zeta_5}{2}
   -\frac{27 \zeta_4}{4}
   \w
    \qquad ~ ~~~~~~
   -\frac{179 \zeta_3}{8}
   +\frac{2075 }{256}
   \Big)\epsilon ^6
   + \ca O(\epsilon^7)\Bigg] + \ca O\Big(\frac 1{n^3}\Big)\ .
\eea
This expansion agrees with Eq.~\eqref{phic} expanded in $1/n$. 
Even though not all 6-loop diagrams contribute to the $1/n^2$ term,   the comparison with the large-$n$ expansion is   a very strong consistency check.
\begin{table}[t]
\vspace*{-2mm}
	\centering	\caption{Numerical values for the exponent $\omega$   in $d=3$. }	\label{tab:omega-numeric}	\begin{tabular}{cccc}
	\toprule
		$\quad n \quad$ & SC & KP17  & other  \\
		\midrule
		-2 & 0.828(13) &0.819(7) &\\
		-1 & 0.86(2) & 0.848(15)&\\
		0 & 0.846(15) & 0.841(13) \cite{KompanietsPanzer2017}& 0.904(5)  \cite{Belohorec1997,ClisbyDunweg2016}\\
		1 &0.827(13) &0.820(7) \cite{KompanietsPanzer2017}& 0.830(2) \cite{El-ShowkPaulosPolandRychkovSimmons-DuffinVichi2014}\\
		2 &0.808(7) &0.804(3) \cite{KompanietsPanzer2017}& 0.811(10) \cite{Castedo-Echeverrivon-HarlingSerone2016}\\
		3 &0.794(4)& 0.795(7) \cite{KompanietsPanzer2017}&0.791(22) \cite{Castedo-Echeverrivon-HarlingSerone2016}\\
		4 & 0.7863(9) & 0.794(9) \cite{KompanietsPanzer2017}&0.817(30) \cite{Castedo-Echeverrivon-HarlingSerone2016}\\
\bottomrule
	\end{tabular}\end{table}

\section{Conclusion and Perspectives}
\label{s:Conclusion}
In this article, we   evaluated the fractal dimension of critical lines in the $O(n)$ model, yielding the fractal dimension of loop-erased random walks ($n=-2$), self-avoiding walks ($n=0$), as well as the propagator line for the Ising model ($n=1$) and the XY model ($n=2$). 
Our predictions from the $
\epsilon$-expansion at 6-loop order are in excellent agreement with numerical simulations in $d=3$,  for the larger values of $n$ even exceeding the numerically obtained precision. 
This was possible through a combination of several resummation techniques, including a self-consistent one introduced here. Analyzing its behavior in dimension $d=2$ to determine the most suitable quantity to be resummed allowed us to   improve the precision for the remaining exponents, especially $\nu$, yielding now an agreement of  $3 \times 10^{-4}$  for the Ising model in $d=3$, as compared to the conformal bootstrap. 

\smallskip

We plan to extend this work in several directions: 
\begin{itemize}
\item
Analyze the analytic structure of the critical exponents as a function of $d$ and $\epsilon$ to better catch the singularities in $d=2$, and thus obtain more precise resummations in $d=3$ for all exponents. 
\item use the 7-loop results of   \cite{Schnetz2018} to improve our estimates. 
\item estimate universal amplitudes appearing in the log-CFT for self-avoiding polymers. 
\end{itemize}


\acknowledgements
It is a pleasure to thank A.A.~Fedorenko for insightful discussions.  
The work of M.K. was supported by the Foundation for the Advancement of Theoretical Physics and Mathematics ``BASIS'' (grant 18-1-2-43-1). M.K.   thanks LPENS for hospitality during the work on this paper. 

%
%

\ifx\doi\undefined
\providecommand{\doi}[2]{\href{http://dx.doi.org/#1}{#2}}
\else
\renewcommand{\doi}[2]{\href{http://dx.doi.org/#1}{#2}}
\fi
\providecommand{\link}[2]{\href{http://#1}{#2}}
\providecommand{\arxiv}[1]{\href{http://arxiv.org/abs/#1}{#1}}


\begin{thebibliography}{10}

\bibitem{ItzyksonDrouffe1}
C.~Itzykson and J.-M. Drouffe,
\newblock \doi{10.1017/CBO9780511622779}{\rm {\em Statistical Field
  Theory}}\null, {\em {\em Volume}~1} of {\em Cambridge Monographs on
  Mathematical Physics},
\newblock Cambridge University Press, 1989.

\bibitem{ItzyksonDrouffe2}
C.~Itzykson and J.-M. Drouffe,
\newblock \doi{10.1017/CBO9780511622786}{\rm {\em Statistical Field
  Theory}}\null, {\em {\em Volume}~2} of {\em Cambridge Monographs on
  Mathematical Physics},
\newblock Cambridge University Press, 1989.

\bibitem{ArisueFujiwara2003}
H.~Arisue and T.~Fujiwara,
\newblock {\em New algorithm of the high-temperature expansion for the {Ising}
  model in three dimensions},
\newblock \doi{https://doi.org/10.1016/S0920-5632(03)01701-8}{\rm Nucl. Phys. B
  Proc. Suppl. {\bf 119} (2003)   855--857}\null.

\bibitem{ArisueFujiwaraTabata2004}
H.~Arisue, T.~Fujiwara  and K.~Tabata,
\newblock {\em Higher orders of the high-temperature expansion for the {Ising}
  model in three dimensions},
\newblock \doi{https://doi.org/10.1016/S0920-5632(03)02709-9}{\rm Nucl. Phys. B
  Proc. Suppl. {\bf 129-130} (2004)   774--776}\null.

\bibitem{ButeraComi1995}
P.~Butera and M.~Comi,
\newblock {\em Extended high-temperature series for the n-vector spin models on
  three-dimensional bipartite lattices},
\newblock \doi{10.1103/PhysRevB.52.6185}{\rm Phys. Rev. B {\bf 52} (1995)
  6185--6188}\null.

\bibitem{ButeraComi2000}
P.~Butera and M.~Comi,
\newblock {\em Extension to order $\ensuremath{\beta}{}^{23}$ of the
  high-temperature expansions for the spin-$\frac{1}{2}$ {Ising} model on
  simple cubic and body-centered cubic lattices},
\newblock \doi{10.1103/PhysRevB.62.14837}{\rm Phys. Rev. B {\bf 62} (2000)
  14837--14843}\null.

\bibitem{ButeraComi1999}
P.~Butera and M.~Comi,
\newblock {\em Critical specific heats of the {$N$}-vector spin models on the
  simple cubic and bcc lattices},
\newblock \doi{10.1103/PhysRevB.60.6749}{\rm Phys. Rev. B {\bf 60} (1999)
  6749--6760}\null.

\bibitem{BatkovichChetyrkinKompaniets2016}
D.V. Batkovich, K.G. Chetyrkin  and M.V. Kompaniets,
\newblock {\em Six loop analytical calculation of the field anomalous dimension
  and the critical exponent $\eta$ in {$O(n)$}-symmetric $\varphi^4$ model},
\newblock \doi{10.1016/j.nuclphysb.2016.03.009}{\rm Nucl. Phys. B {\bf 906}
  (2016)   147}\null,
\newblock \arxiv{arXiv:1601.01960}.

\bibitem{KompanietsPanzer:LL2016}
M.V. Kompaniets and E.~Panzer,
\newblock in Proceedings of {\em
  \href{http://pos.sissa.it/cgi-bin/reader/conf.cgi?confid=260}{Loops and Legs
  in Quantum Field Theory}}, 
  Sissa Medialab, 2016,
\newblock \arxiv{arXiv:1606.09210}.

\bibitem{KompanietsPanzer2017}
M.V. Kompaniets and E.~Panzer,
\newblock {\em Minimally subtracted six-loop renormalization of
  {$O(n)$}-symmetric ${\ensuremath{\phi}}^{4}$ theory and critical exponents},
\newblock \doi{10.1103/PhysRevD.96.036016}{\rm Phys. Rev. D {\bf 96} (2017)
  036016}\null,
\newblock \arxiv{arXiv:1705.06483}.

\bibitem{Amit}
D.J. Amit and V.~Martin-Mayor,
\newblock \doi{10.1142/9789812775313_0014}{\rm {\em Field Theory, the
  Renormalization Group, and Critical Phenomena}}\null,
\newblock World Scientific, Singapore, 3rd edition, 1984.

\bibitem{Zinn}
J.~Zinn-Justin,
\newblock \doi{10.1093/acprof:oso/9780199227198.001.0001}{\rm {\em Quantum
  Field Theory and Critical Phenomena}}\null,
\newblock Oxford University Press, Oxford, 1989.

\bibitem{Vasilev2004}
A.N. Vasil'ev,
\newblock \doi{10.1201/9780203483565}{\rm {\em The Field Theoretic
  Renormalization Group in Critical Behavior Theory and Stochastic
  Dynamics}}\null,
\newblock Chapman \& Hall/CRC, New York, 2004.

\bibitem{ChetyrkinKataevTkachov1981}
K.G. {Chetyrkin}, A.L. {Kataev}  and F.V. {Tkachov},
\newblock {\em Five-loop calculations in the {$g\varphi^4$} model and the
  critical index {$\eta$}},
\newblock \doi{10.1016/0370-2693(81)90968-0}{\rm Phys. Lett. B {\bf 99} (1981)
   147--150}\null.

\bibitem{ChetyrkinKataevTkachov1981b}
K.~G. {Chetyrkin}, A.~L. {Kataev}  and F.~V. {Tkachov},
\newblock {\em Errata},
\newblock \doi{10.1016/0370-2693(81)90176-3}{\rm Phys. Lett. B {\bf 101} (1981)
    457--458}\null.

\bibitem{ChetyrkinGorishnyLarinTkachov1983}
K.~G. {Chetyrkin}, S.~G. {Gorishny}, S.~A. {Larin}  and F.~V. {Tkachov},
\newblock {\em Five-loop renormalization group calculations in the {$g
  \varphi^4$} theory},
\newblock \doi{10.1016/0370-2693(83)90324-6}{\rm Phys. Lett. B {\bf 132} (1983)
    351--354}\null.

\bibitem{Kazakov1983}
D.~I. {Kazakov},
\newblock {\em The method of uniqueness, a new powerful technique for multiloop
  calculations},
\newblock \doi{10.1016/0370-2693(83)90816-X}{\rm Phys. Lett. B {\bf 133} (1983)
    406--410}\null.

\bibitem{KleinertNeuSchulte-FrohlindeChetyrkinLarin1991}
H.~Kleinert, J.~Neu, V.~Schulte-Frohlinde, K.G. Chetyrkin  and S.A. Larin,
\newblock {\em Five-loop renormalization group functions of {$O(n)$}-symmetric
  {$\phi^4$}-theory and {$\epsilon$}-expansions of critical exponents up to
  {$\epsilon^5$}},
\newblock \doi{10.1016/0370-2693(91)91009-K}{\rm Phys. Lett. B {\bf 272} (1991)
    39--44}\null,
\newblock \arxiv{hep-th/9503230}.

\bibitem{Schnetz2018}
O.~Schnetz,
\newblock {\em Numbers and functions in quantum field theory},
\newblock \doi{10.1103/PhysRevD.97.085018}{\rm Phys. Rev. D {\bf 97} (2018)
  085018}\null.

\bibitem{ParisiBook}
G.~Parisi,
\newblock {\em Statistical Field Theory},
\newblock Frontiers in Physics,
\newblock Addison-Wesley, 1988.

\bibitem{BakerNickelGreenMeiron1976}
G.A. Baker, B.G. Nickel, M.S. Green  and D.I. Meiron,
\newblock {\em Ising-model critical indices in three dimensions from the
  {Callan-Symanzik} equation},
\newblock \doi{10.1103/PhysRevLett.36.1351}{\rm Phys. Rev. Lett. {\bf 36}
  (1976)   1351--1354}\null.

\bibitem{GuidaZinn-Justin1998}
R.~Guida and J.~Zinn-Justin,
\newblock {\em Critical exponts of the {$N$}-vector model},
\newblock \doi{10.1088/0305-4470/31/40/006}{\rm J.Phys. A {\bf 31} (1998)
  8103}\null,
\newblock \arxiv{cond-mat/9803240}.

\bibitem{Clisby2017}
N.~{Clisby},
\newblock {\em Scale-free {M}onte {C}arlo method for calculating the critical
  exponent {$\gamma$} of self-avoiding walks},
\newblock \doi{10.1088/1751-8121/aa7231}{\rm J. Phys. A {\bf 50} (2017)
  264003}\null,
\newblock \arxiv{arXiv:1701.08415}.

\bibitem{ClisbyDunweg2016}
N.~{Clisby} and B.~{D{\"u}nweg},
\newblock {\em High-precision estimate of the hydrodynamic radius for
  self-avoiding walks},
\newblock \doi{10.1103/PhysRevE.94.052102}{\rm Phys. Rev. E {\bf 94} (2016)
  052102}\null.

\bibitem{CampostriniHasenbuschPelissettoVicari2006}
M.~{Campostrini}, M.~{Hasenbusch}, A.~{Pelissetto}  and E.~{Vicari},
\newblock {\em Theoretical estimates of the critical exponents of the
  superfluid transition in {$^{4}\mathrm{He}$} by lattice methods},
\newblock \doi{10.1103/PhysRevB.74.144506}{\rm Phys. Rev. B {\bf 74} (2006)
  144506}\null,
\newblock \arxiv{cond-mat/0605083}.

\bibitem{HasenbuschVicari2011}
M.~{Hasenbusch} and E.~{Vicari},
\newblock {\em Anisotropic perturbations in three-dimensional
  {$O(N)$}-symmetric vector models},
\newblock \doi{10.1103/PhysRevB.84.125136}{\rm Phys. Rev. B {\bf 84} (2011)
  125136}\null,
\newblock \arxiv{1108.0491}.

\bibitem{Deng2006}
Y.~Deng,
\newblock {\em Bulk and surface phase transitions in the three-dimensional
  {$O(4)$} spin model},
\newblock \doi{10.1103/PhysRevE.73.056116}{\rm Phys. Rev. E {\bf 73} (2006)
  056116}\null.

\bibitem{Nienhuis1987}
B.~Nienhuis,
\newblock {\em Coulomb Gas Formulation of Two-dimensional Phase Transitions},
  {\em {\em Volume}~11} of {\em Phase Transitions and Critical Phenomena},
\newblock Academic Press, London, 1987.

\bibitem{HenkelCFT}
M.~Henkel,
\newblock \doi{10.1007/978-3-662-03937-3}{\rm {\em Conformal Invariance and
  Critical Phenomena}}\null,
\newblock Springer, Berlin, Heidelberg, 1999.

\bibitem{DiFrancescoMathieuSenechal}
P.~Di Francesco, P.~Mathieu  and D.~S\'en\'echal,
\newblock \doi{10.1007/978-1-4612-2256-9}{\rm {\em Conformal Field
  Theory}}\null,
\newblock Springer, New York, 1997.

\bibitem{NishimoriOrtiz2011}
H.~Nishimori and G.~Ortiz,
\newblock \doi{10.1093/acprof:oso/9780199577224.001.0001}{\rm {\em Elements of
  phase transitions and critical phenomena}}\null,
\newblock Oxford University Press, 2011.

\bibitem{PolandRychkovVichi2019}
D.~Poland, S.~Rychkov  and A.~Vichi,
\newblock {\em The conformal bootstrap: Theory, numerical techniques, and
  applications},
\newblock \doi{10.1103/RevModPhys.91.015002}{\rm Rev. Mod. Phys. {\bf 91}
  (2019)   015002}\null.

\bibitem{KosPolandSimmons-DuffinVichi2016}
F.~{Kos}, D.~{Poland}, D.~{Simmons-Duffin}  and A.~{Vichi},
\newblock {\em Precision islands in the {Ising} and {$O(N)$} models},
\newblock \doi{10.1007/JHEP08(2016)036}{\rm JHEP {\bf 08} (2016)   036}\null,
\newblock \arxiv{arXiv:1603.04436}.

\bibitem{El-ShowkPaulosPolandRychkovSimmons-DuffinVichi2014}
S.~{El-Showk}, M.~F. {Paulos}, D.~{Poland}, S.~{Rychkov}, D.~Simmons-Duffin
  and A.~{Vichi},
\newblock {\em Solving the 3d {Ising} model with the conformal bootstrap ii.
  {$c$}-minimization and precise critical exponents},
\newblock \doi{10.1007/s10955-014-1042-7}{\rm J. Stat. Phys. {\bf 157} (2014)
  869--914}\null,
\newblock \arxiv{arXiv:1403.4545}.

\bibitem{Castedo-Echeverrivon-HarlingSerone2016}
A.~{Castedo Echeverri}, B.~{von Harling}  and M.~{Serone},
\newblock {\em The effective bootstrap},
\newblock \doi{10.1007/JHEP09(2016)097}{\rm JHEP {\bf 09} (2016)   097}\null,
\newblock \arxiv{arXiv:1606.02771}.

\bibitem{GuttmanInDombLebowitz13}
A.J. Guttmann.
\newblock {\em {\em Volume}~13} of {\em Phase Transitions and Critical
  Phenomena},
\newblock Academic Press, London, 1987.

\bibitem{Zinn-Justin2001}
J.~Zinn-Justin,
\newblock {\em Precise determination of critical exponents and equation of
  state by field theory methods},
\newblock \doi{10.1016/S0370-1573(00)00126-5}{\rm Phys. Rep. {\bf 344} (2001)
  159--178}\null.

\bibitem{HooftVeltman1972}
G.~'t~Hooft and M.~Veltman,
\newblock {\em Regularization and renormalization of gauge fields},
\newblock \doi{https://doi.org/10.1016/0550-3213(72)90279-9}{\rm Nucl. Phys. B
  {\bf 44} (1972)   189 -- 213}\null.

\bibitem{Lawler1980}
G.F. Lawler,
\newblock {\em A self-avoiding random walk},
\newblock \doi{10.1215/S0012-7094-80-04741-9}{\rm Duke Math. J. {\bf 47} (1980)
    655--693}\null.

\bibitem{Falconer1986}
K.J. Falconer,
\newblock \doi{10.1017/CBO9780511623738}{\rm {\em The Geometry of Fractal
  Sets}}\null,
\newblock Cambridge University Press, Cambridge, U.K., 1986.

\bibitem{LawlerSchrammWerner2004}
G.F. Lawler, O.~Schramm  and W.~Werner,
\newblock {\em Conformal invariance of planar loop-erased random walks and
  uniform spanning trees},
\newblock \doi{10.1007/978-1-4419-9675-6_30}{\rm Ann. Probab. {\bf 32} (2004)
  939--995}\null,
\newblock \arxiv{arXiv:math/0112234}.


\bibitem{DeGennes1972}
P.-G.~De Gennes,
\newblock {\em Exponents for the excluded volume problem as derived by the
  {Wilson} method},
\newblock \doi{10.1016/0375-9601(72)90149-1}{\rm Phys. Lett. {\bf A 38} (1972)
   339--340}\null.


\bibitem{WieseFedorenko2018}
{K.J.} Wiese and A.A. Fedorenko,
\newblock {\em Field theories for loop-erased random walks},
\newblock \doi{10.1016/j.nuclphysb.2019.114696}{\rm Nucl. Phys. B {\bf 946}
  (2019)   114696}\null,
\newblock \arxiv{arXiv:1802.08830}.

\bibitem{WieseFedorenko2019}
K.J. Wiese and A.A. Fedorenko,
\newblock {\em Depinning transition of charge-density waves: Mapping onto
  ${O}(n)$ symmetric $\phi^4$ theory with $n\to -2$ and loop-erased random
  walks},
\newblock \doi{10.1103/PhysRevLett.123.197601}{\rm Phys. Rev. Lett. {\bf 123}
  (2019)   197601}\null,
\newblock \arxiv{arXiv:1908.11721}.



\bibitem{Wilson2010}
D.B. Wilson,
\newblock {\em Dimension of the loop-erased random walk in three dimensions},
\newblock \doi{10.1103/PhysRevE.82.062102}{\rm Phys. Rev. E {\bf 82} (2010)
  062102}\null,
\newblock \arxiv{arXiv:1008.1147}.

\bibitem{WinterJankeSchakel2008}
F.~Winter, W.~Janke  and A.M.J. Schakel,
\newblock {\em Geometric properties of the three-dimensional {Ising} and {$XY$}
  models},
\newblock \doi{10.1103/PhysRevE.77.061108}{\rm Phys. Rev. E {\bf 77} (2008)
  061108}\null,
\newblock \arxiv{arXiv:0803.2177}.

\bibitem{ProkofevSvistunov2006}
N.~Prokof'ev and B.~Svistunov,
\newblock {\em Comment on ``{Hausdorff} dimension of critical fluctuations in
  {Abelian} gauge theories''},
\newblock \doi{10.1103/PhysRevLett.96.219701}{\rm Phys. Rev. Lett. {\bf 96}
  (2006)   219701}\null.

\bibitem{Kirkham1981}
J.E. Kirkham,
\newblock {\em Calculation of crossover exponent from {Heisenberg} to {Ising}
  behaviour using the fourth-order $\epsilon$ expansion},
\newblock \doi{10.1088/0305-4470/14/11/004}{\rm J. Phys. A {\bf 14} (1981)
  L437--L442}\null.

\bibitem{Gracey2002}
J.~A. Gracey,
\newblock {\em Crossover exponent in $\mathrm{O}({N})$
  ${\ensuremath{\varphi}}^{4}$ theory at ${O(1/N}^{2})$},
\newblock \doi{10.1103/PhysRevE.66.027102}{\rm Phys. Rev. E {\bf 66} (2002)
  027102}\null.

\bibitem{ShimadaHikami2016}
H.~Shimada and S.~Hikami,
\newblock {\em Fractal dimensions of self-avoiding walks and {Ising}
  high-temperature graphs in 3d conformal bootstrap},
\newblock \doi{10.1007/s10955-016-1658-x}{\rm J. Stat. Phys {\bf 165} (2016)
  1006--1035}\null,
\newblock \arxiv{arXiv:1509.04039}.

\bibitem{Kleinert2000}
H.~Kleinert,
\newblock {\em Variational resummation for $\epsilon$-expansions of critical
  exponents of nonlinear ${O}(n)$-symmetric $\sigma$-model in $2+\epsilon$
  dimensions},
\newblock \doi{10.1016/S0375-9601(99)00833-6}{\rm Phys. Lett. A {\bf 264}
  (2000)   357--365}\null.

\bibitem{KiskisNarayananVranas1993}
J.~Kiskis, R.~Narayanan  and P.~Vranas,
\newblock {\em The {Hausdorff} dimension of random walks and the correlation
  length critical exponent in {Euclidean} field theory},
\newblock \doi{10.1007/BF01054349}{\rm J. Stat. Phys. {\bf 73} (1993)
  765--774}\null.

\bibitem{RohrerGerber1977}
H.~Rohrer and Ch. Gerber,
\newblock {\em Bicritical and tetracritical behavior of
  {${\mathrm{GdAlO}_{3}}$}},
\newblock \doi{10.1103/PhysRevLett.38.909}{\rm Phys. Rev. Lett. {\bf 38} (1977)
    909--912}\null.

\bibitem{Domann1979}
G.~Domann,
\newblock {\em Optical measurements on {$\mathrm{TbPO}_4$} near the bicritical
  point},
\newblock \doi{10.1016/0304-8853(79)90057-X}{\rm J. Magn. Magn. Mater. {\bf 13}
  (1979)   163--166}\null.

\bibitem{MajkrzakAxeBruce1980}
C.F. Majkrzak, J.D. Axe  and A.D. Bruce,
\newblock {\em Critical behavior at the incommensurate structural phase
  transition in {${\mathrm{K}}_{2}$\rm Se${\mathrm{O}}_{4}$}},
\newblock \doi{10.1103/PhysRevB.22.5278}{\rm Phys. Rev. B {\bf 22} (1980)
  5278--5283}\null.

\bibitem{WalischPerez-MatoPetersson1989}
R.~Walisch, J.~M. Perez-Mato  and J.~Petersson,
\newblock {\em {NMR} determination of the nonclassical critical exponents
  ${\beta}$ and ${\bar\beta}$ in incommensurate
  {${\mathrm{Rb}}_{2}$${\mathrm{ZnCl}}_{4}$}},
\newblock \doi{10.1103/PhysRevB.40.10747}{\rm Phys. Rev. B {\bf 40} (1989)
  10747--10752}\null.

\bibitem{WuYoungShaoGarlandBirgeneauHeppke1994}
Lei Wu, M.J. Young, Y.~Shao, C.W. Garland, R.J. Birgeneau  and G.~Heppke,
\newblock {\em Critical behavior of the second harmonic in a density wave
  system},
\newblock \doi{10.1103/PhysRevLett.72.376}{\rm Phys. Rev. Lett. {\bf 72} (1994)
    376--379}\null.

\bibitem{ShapiraOliveira1978}
Y.~Shapira and N.~F. Oliveira,
\newblock {\em Crossover behavior of the magnetic phase boundary of the
  isotropic antiferromagnet {$\mathrm{RbMnF}_{3}$} from ultrasonic
  measurements},
\newblock \doi{10.1103/PhysRevB.17.4432}{\rm Phys. Rev. B {\bf 17} (1978)
  4432--4443}\null.

\bibitem{KingRohrer1979}
A.~R. King and H.~Rohrer,
\newblock {\em Spin-flop bicritical point in {${\mathrm{MnF}}_{2}$}},
\newblock \doi{10.1103/PhysRevB.19.5864}{\rm Phys. Rev. B {\bf 19} (1979)
  5864--5876}\null.

\bibitem{Zhang1997}
S.-C. Zhang,
\newblock {\em A unified theory based on {SO(5)} symmetry of superconductivity
  and antiferromagnetism},
\newblock \doi{10.1126/science.275.5303.1089}{\rm Science {\bf 275} (1997)
  1089--1096}\null.

\bibitem{Hu2001}
X.~Hu,
\newblock {\em Bicritical and tetracritical phenomena and scaling properties of
  the {SO(5)} theory},
\newblock \doi{10.1103/PhysRevLett.87.057004}{\rm Phys. Rev. Lett. {\bf 87}
  (2001)   057004}\null.

\bibitem{Hu2002}
X.~Hu,
\newblock {\em Hu replies},
\newblock \doi{10.1103/PhysRevLett.88.059704}{\rm Phys. Rev. Lett. {\bf 88}
  (2002)   059704}\null.

\bibitem{Aharony2002}
A.~Aharony,
\newblock {\em Comment on ``{Bicritical} and tetracritical phenomena and
  scaling properties of the {SO(5)} theory''},
\newblock \doi{10.1103/PhysRevLett.88.059703}{\rm Phys. Rev. Lett. {\bf 88}
  (2002)   059703}\null.

\bibitem{PfeutyJasnowFisher1974}
P.~Pfeuty, D.~Jasnow  and M.E. Fisher,
\newblock {\em Crossover scaling functions for exchange anisotropy},
\newblock \doi{10.1103/PhysRevB.10.2088}{\rm Phys. Rev. B {\bf 10} (1974)
  2088--2112}\null.

\bibitem{CalabresePelissettoVicari2002}
P.~Calabrese, A.~Pelissetto  and E.~Vicari,
\newblock {\em Critical structure factors of bilinear fields in
  $\mathrm{O}({N})$ vector models},
\newblock \doi{10.1103/PhysRevE.65.046115}{\rm Phys. Rev. E {\bf 65} (2002)
  046115}\null.

\bibitem{EichhornMesterhazyScherer2013}
A.~Eichhorn, D.~Mesterh\'azy  and M.M. Scherer,
\newblock {\em Multicritical behavior in models with two competing order
  parameters},
\newblock \doi{10.1103/PhysRevE.88.042141}{\rm Phys. Rev. E {\bf 88} (2013)
  042141}\null.

\bibitem{SapozhnikovShiraishi2018}
A.~Sapozhnikov and D.~Shiraishi,
\newblock {\em On {Brownian} motion, simple paths, and loops},
\newblock \doi{10.1007/s00440-017-0817-6}{\rm Probab. Th. Rel. Fields {\bf 172}
  (2018)   615--662}\null,
\newblock \arxiv{arXiv:1512.04864}.

\bibitem{Nienhuis1982}
B.~Nienhuis,
\newblock {\em Exact critical point and critical exponents of $\mathrm{O}(n)$
  models in two dimensions},
\newblock \doi{10.1103/PhysRevLett.49.1062}{\rm Phys. Rev. Lett. {\bf 49}
  (1982)   1062--1065}\null.

\bibitem{Duplantier1992}
B.~Duplantier,
\newblock {\em Loop-erased self-avoiding walks in two dimensions: exact
  critical exponents and winding numbers},
\newblock \doi{10.1016/0378-4371(92)90575-B}{\rm Physica A {\bf 191} (1992)
  516--522}\null.

\bibitem{LyklemaEvertszPietronero1986}
J.W. Lyklema, C.~Evertsz  and L.~Pietronero,
\newblock {\em The {Laplacian} random walk},
\newblock \doi{0295-5075/2/i=2/a=001}{\rm EPL {\bf 2} (1986)  ~77}\null.

\bibitem{Lawler2006}
G.F. Lawler,
\newblock {\em The {Laplacian}-$b$ random walk and the {Schramm-Loewner}
  evolution},
\newblock Illinois J. Math. {\bf 50} (2006)   701--746.

\bibitem{Cardy2005}
J.~Cardy,
\newblock {\em {SLE} for theoretical physicists},
\newblock \doi{10.1016/j.aop.2005.04.001}{\rm Annals of Physics {\bf 318}
  (2005)   81--118}\null,
\newblock \arxiv{cond-mat/0503313v2}.

\bibitem{BelavinPolyakovZamolodchikov1984}
A.A. Belavin, A.M. Polyakov  and A.B. Zamolodchikov,
\newblock {\em Infinite conformal symmetry in two-dimensional quantum field
  theory},
\newblock \doi{10.1016/0550-3213(84)90052-X}{\rm Nucl. Phys. B {\bf 241} (1984)
    333--380}\null.
    
    
\bibitem{DotsenkoCFT}
Vl.~S. Dotsenko,
\newblock {\em \link{cel.archives-ouvertes.fr/cel-00092929}{S\'erie de cours
  sur la th\'eorie conforme}},
\newblock Lecture notes, Universit\'es Paris VI and VII.


\bibitem{JankeSchakel2010}
W.~Janke and A.M.J. Schakel,
\newblock {\em Holographic interpretation of two-dimensional {$O(N)$} models
  coupled to quantum gravity},
\newblock (2010),
\newblock \arxiv{arXiv:1003.2878}.


\bibitem{RushkinBettelheimGruzbergWiegmann2007}
I.~Rushkin, E.~Bettelheim, I.A. Gruzberg  and P.~Wiegmann,
\newblock {\em Critical curves in conformally invariant statistical systems},
\newblock \doi{10.1088/1751-8113/40/9/020}{\rm J. Phys. A {\bf 40} (2007)
  2165--2195}\null,
\newblock \arxiv{cond-mat/0610550}.



\bibitem{BloeteKnopsNienhuis1992}
H.W.J. Bl\"ote, Y.M.M. Knops  and B.~Nienhuis,
\newblock {\em Geometrical aspects of critical {Ising} configurations in 2
  dimensions},
\newblock \doi{10.1103/PhysRevLett.68.3440}{\rm Phys. Rev. Lett. {\bf 68}
  (1992)   3440--3443}\null.

\bibitem{BarouchMcCoyWu1973}
E.~Barouch, B.M. McCoy  and T.T. Wu,
\newblock {\em Zero-field susceptibility of the two-dimensional ising model
  near ${T}_{c}$},
\newblock \doi{10.1103/PhysRevLett.31.1409}{\rm Phys. Rev. Lett. {\bf 31}
  (1973)   1409--1411}\null.

\bibitem{CalabreseCaselleCeliPelissettoVicari2000}
P.~Calabrese, M.~Caselle, A.~Celi, A.~Pelissetto  and E.~Vicari,
\newblock {\em Non-analyticity of the {Callan-Symanzik} $\beta$-function of
  two-dimensional {$O(N)$} models},
\newblock \doi{10.1088/0305-4470/33/46/301}{\rm J. Phys. A {\bf 33} (2000)
  8155--8170}\null.

\bibitem{Cardy2018}
J.~Cardy,
\newblock {\em ${T\overline T}$ deformations of non-{Lorentz} invariant field
  theories},
\newblock \doi{10.1007/JHEP10(2018)186}{\rm JHEP {\bf 10} (2018)   186}\null,
\newblock \arxiv{arXiv:1809.07849}.

\bibitem{Belohorec1997}
P.~Belohorec,
\newblock {\em Renormalization group calculation of the universal critical
  exponents of a polymer molecule},
\newblock
  \link{http://www.collectionscanada.gc.ca/obj/s4/f2/dsk3/ftp04/nq24397.pdf}{\rm
  PhD thesis, University of Guelph, Ontario, Canada}\null, 1997.

\bibitem{CampostriniHasenbuschPelissettoRossiVicari2002}
M.~{Campostrini}, M.~{Hasenbusch}, A.~{Pelissetto}, P.~{Rossi}  and
  E.~{Vicari},
\newblock {\em Critical exponents and equation of state of the
  three-dimensional {Heisenberg} universality class},
\newblock \doi{10.1103/PhysRevB.65.144520}{\rm Phys. Rev. B {\bf 65} (2002)
  144520}\null,
\newblock \arxiv{cond-mat/0110336}.

\end{thebibliography}
\end{document}